\documentclass[11pt,a4paper]{article}
\pdfoutput=1
\usepackage{jstyle}
\usepackage{amsmath, amsfonts, amssymb}
\usepackage{graphicx, hyperref, color, verbatim}
\usepackage[dvipsnames]{xcolor}
\usepackage{float}
\usepackage{caption}
\usepackage{subcaption}

\makeatletter
\DeclareFontFamily{OMX}{MnSymbolE}{}
\DeclareSymbolFont{MnLargeSymbols}{OMX}{MnSymbolE}{m}{n}
\SetSymbolFont{MnLargeSymbols}{bold}{OMX}{MnSymbolE}{b}{n}
\DeclareFontShape{OMX}{MnSymbolE}{m}{n}{
    <-6>  MnSymbolE5
   <6-7>  MnSymbolE6
   <7-8>  MnSymbolE7
   <8-9>  MnSymbolE8
   <9-10> MnSymbolE9
  <10-12> MnSymbolE10
  <12->   MnSymbolE12
}{}
\DeclareFontShape{OMX}{MnSymbolE}{b}{n}{
    <-6>  MnSymbolE-Bold5
   <6-7>  MnSymbolE-Bold6
   <7-8>  MnSymbolE-Bold7
   <8-9>  MnSymbolE-Bold8
   <9-10> MnSymbolE-Bold9
  <10-12> MnSymbolE-Bold10
  <12->   MnSymbolE-Bold12
}{}

\let\llangle\@undefined
\let\rrangle\@undefined
\DeclareMathDelimiter{\llangle}{\mathopen}%
                     {MnLargeSymbols}{'164}{MnLargeSymbols}{'164}
\DeclareMathDelimiter{\rrangle}{\mathclose}%
                     {MnLargeSymbols}{'171}{MnLargeSymbols}{'171}
\makeatother

\title{Flat-space limit of defect correlators and stringy AdS form factors}

\author[a]{Luis F. Alday,}
\author[b,c]{Xinan Zhou}

\affiliation[a]{Mathematical Institute, University of Oxford, Andrew Wiles Building, Radcliffe Observatory Quarter, Woodstock Road, Oxford, OX2 6GG, U.K.}
\affiliation[b]{Kavli Institute for Theoretical Sciences, University of Chinese Academy of Sciences, Beijing 100190, China}
\affiliation[c]{School of Physical Science and Technology, ShanghaiTech University, Shanghai 201210, China.}

\emailAdd{alday@maths.ox.ac.uk}
\emailAdd{xinan.zhou@ucas.ac.cn}

\abstract{We study AdS form factors, given by the Mellin representation for CFT correlators of local operators in the presence of extended defects. We propose a formula for taking (and expanding around) the flat-space limit. This formula relates the flat-space form factors for particles scattering off an extended object to the high-energy limit of the Mellin amplitude, via a Borel transform. We check the validity of our proposal in a number of examples. As an application, we study the two-point function of local operators in the presence of a 't Hooft loop in 4d $\mathcal{N}=4$ SYM, and compute the first few orders of stringy corrections to the AdS form factor of gravitons scattering off a D1 brane. }

\begin{document}
\maketitle
\tableofcontents

\newpage

\section{Introduction}
Holographic correlators are the most fundamental observables for exploring and exploiting the AdS/CFT correspondence. Traditionally,  these observables are computed by the diagrammatic expansion method in AdS. Unfortunately, this method quickly meets unsurmountable technical difficulties beyond just the few simplest examples, which resulted in almost a decade long hiatus in the study of holographic correlators. However, recent years have witnessed significant progress following the introduction of the bootstrap strategy \cite{Rastelli:2016nze,Rastelli:2017udc}. This  strategy relies on the principles of symmetry and consistency conditions and allows us to circumvent the difficulties of the traditional method. A plethora of new results were obtained using this strategy (see \cite{Bissi:2022mrs} for a review). For example, all infinitely many tree-level four-point functions are known in Mellin space \cite{Mack:2009mi,Penedones:2010ue} for supergravities in maximally superconformal backgrounds \cite{Rastelli:2016nze,Rastelli:2017udc,Alday:2020lbp,Alday:2020dtb}, as well as SYM in half maximally superconformal backgrounds \cite{Alday:2021odx} -- a feat which is not imaginable using the traditional method. On the other hand, it is also interesting to extend the bootstrap strategy beyond supergravity and SYM, and study the rich physics and mathematics encoded in the corrections from string theory and M-theory which are manifested as higher-derivative contact interactions. The study of these corrections was initiated in \cite{Alday:2014tsa,Goncalves:2014ffa,Chester:2018aca,Binder:2018yvd} and two new ingredients are important. The first is the flat-space limit formula which relates the flat-space amplitude to the high-energy limit of the Mellin amplitude \cite{Penedones:2010ue}
\begin{equation}\label{flatspacePenedonesintro}
\mathcal{M}(s_{ij})\propto \int_0^\infty d\beta \beta^{\frac{1}{2}\sum_i\Delta_i-\frac{d}{2}-1}e^{-\beta}\mathcal{A}\left(S_{ij}=\frac{2\beta}{R^2}s_{ij}\right)\;,\quad R\to\infty\;.
\end{equation}
This formula, featuring a Borel transformation, allows us to fix all the leading terms in the corrections which survive the flat-space limit by comparing with the known string amplitudes in flat space. The second is the input from supersymmetric localization, in particular in the form of  integrated correlators \cite{Binder:2018yvd,Binder:2019jwn,Chester:2020dja}. They are related to derivatives of the partition functions and can be computed exactly by using localization techniques. Using these localization constraints, one can also probe terms which are subleading in the flat-space limit and fix the stringy corrections up to a certain derivative order. Beyond this, localization is not enough, because at each order localization only gives a small number of constraints while the number of independent superconformal contact corrections grows indefinitely with the number of derivatives. This obstacle was overcome in \cite{Alday:2022uxp,Alday:2022xwz,Alday:2023jdk,Alday:2023mvu} where intuition from a putative world-sheet was used in combination with single-valuedness to fix infinitely many correction coefficients and resum them into AdS string amplitudes in an expansion with respect to the inverse AdS radius $1/R$. In this approach the flat-space formula (\ref{flatspacePenedonesintro}) also plays a pivotal role. It was proposed that for finite $R$ ({i.e., not only $R\to\infty$) this formula gives a {\it working definition} for string amplitudes in AdS \cite{Alday:2023mvu}. Note that the naive resummation of higher-derivative corrections in the Mellin amplitude is divergent even at the leading order in $1/R$ (corresponding to Virasoro-Shapiro amplitude in flat space) and only the Borel transformed series is convergent. 

These results are for CFTs in infinite empty flat space. It is interesting to consider CFTs with conformal defects and interfaces (or boundaries), which we will collectively refer to as defects in what follows. They are important because of not only their experimental relevance but also diverse formal applications in Quantum Field Theory and string theory.  In the simplest case, the defects are dual to probe branes in AdS space. While holographic correlators in defect-free CFTs are dual to on-shell scattering amplitudes in AdS, for CFT correlators in the presence of a defect we encounter new observables. These holographic correlators are dual to {\it form factors} in AdS describing the scattering of particles (or strings) with an extended object.\footnote{Here we are considering holographic correlators where the operators are inserted away from the defects. The case where all operators are inserted on the defect has been considered in, e.g., \cite{Giombi:2017cqn,Drukker:2020swu,Ferrero:2021bsb,Giombi:2023zte,Ferrero:2023znz,Ferrero:2023gnu}, and are dual to scattering amplitudes in a restricted AdS subspace.} Many of the tools available for defect-free CFTs have been extended to the case of CFTs with defects. For instance, a defect version of Mellin space has been developed in \cite{Rastelli:2017ecj,Goncalves:2018fwx}. At tree-level supergravity, a bootstrap strategy extending \cite{Rastelli:2016nze,Rastelli:2017udc} has also been initiated in \cite{Gimenez-Grau:2023fcy} and applied to the case of two-point functions with a $\frac{1}{2}$-BPS Wilson loop in 4d $\mathcal{N}=4$ SYM. Furthermore, a defect version of the unitarity method in AdS, generalizing the defect-free one \cite{Aharony:2016dwx}, was developed in \cite{Chen:2024orp}. This method was used to obtain the first one-loop two-point function with a surface defect in 6d $\mathcal{N}=(2,0)$ theory which used the tree-level results \cite{Chen:2023yvw} as an input. On the field theory side, the localization calculation involving two mass derivatives of the mass-deformed partition function has been systematically analyzed in \cite{Pufu:2023vwo} for general Wilson-'t Hooft loops in $\mathcal{N}=4$ SYM. The precise identification of these mass derivatives with integrated correlators was made in \cite{Billo:2023ncz,Dempsey:2024vkf,Billo:2024kri}. Moreover, the flat-space string theory form factor for gravitons scattering off D-branes is also known \cite{Klebanov:1995ni,Hashimoto:1996bf,Garousi:1996ad}. With these results, we are almost in a position to launch a parallel program to study systematically the stringy corrections to holographic defect correlators. The only missing piece is a flat-space limit formula analogous to (\ref{flatspacePenedonesintro}), which will allow us to bridge together all these ingredients. In this paper, we will close this gap. We find that the flat-space form factor (focusing on two-point functions for simplicity) can be similarly extracted from Mellin space in the high-energy limit via a Borel transform
\begin{equation}
\mathcal{M}(\delta,\gamma)\propto \int_0^\infty d\beta \beta^{\frac{1}{2}(\Delta_1+\Delta_2)-\frac{p}{2}-1}e^{-\beta} \mathcal{A}\left(S=-\frac{2\delta\beta}{R^2},Q=\frac{2\gamma\beta}{R^2}\right)\;,\quad R\to\infty\;.
\end{equation}
This is a central result of this paper. It would be very interesting to derive this formula as was done in  \cite{Penedones:2010ue}.  In this paper we will instead verify the validity of this formula in a number of nontrivial examples, both at the tree level as well as at the one-loop level. 

This flat-space formula has many nontrivial implications. First, for finite $R$ it gives a working definition of string  form factors in AdS. It also makes it possible to initiate precision bootstrap studies of stringy effects in holographic correlators in various theories when defects are present.  As a simple demonstration, in this paper we will use it to study the stringy corrections in the two-point function of $\frac{1}{2}$-BPS operators with a  't Hooft loop in $\mathcal{N}=4$ SYM. Via holography, the 't Hooft loop is dual to a D1 brane which occupies an $AdS_2$ subspace of $AdS_5\times S^5$. As we will explain in detail, the stringy corrections to the two-point function have the following structure when expanded around the large 't Hooft coupling $\lambda$
\begin{equation}
\mathcal{M}=\mathcal{M}_{\rm sugra} f(\lambda)+\lambda^{-1}\mathcal{M}^{\rm h.d.}_{L=2}+\lambda^{-\frac{3}{2}}\mathcal{M}^{\rm h.d.}_{L=3}+\lambda^{-2}\mathcal{M}^{\rm h.d.}_{L=4}+\ldots\;.
\end{equation}
The first term corresponds to the supergravity solution $\mathcal{M}_{\rm sugra}$. This is multiplied by a nontrivial function $f(\lambda)$, which arises from the coupling dependence in the one-point function and can be computed exactly by using localization. In addition, we have an infinite tower of higher-derivative contact interactions which can be seen as arising from integrating out massive string states. Using our flat-space limit formula together with the constraints from localization, we will fix all stringy corrections to the Mellin amplitude with eight derivatives or lower, up to a linear combination of coefficients in the highest eight-derivative part $\mathcal{M}^{\rm h.d.}_{L=4}$. 

The rest of this paper is organized as follows. In Section \ref{Sec:defectfreeflatspace}, we review the defect-free case and see how the flat-space formula can be obtained by analyzing the scalar exchange Witten diagram. We apply this strategy in Section \ref{Sec:flatspaceformuladefect} to defect CFTs and derive the defect version of the flat-space limit formula. We then perform several nontrivial checks in Section \ref{Sec:checks}. We consider the two-point function in $\mathcal{N}=4$ SYM with a 't Hooft loop in Section \ref{Sec:stringycorrections}. We start with a review of kinematics in Section \ref{Sec:twistedandintegrated} which includes also the definitions of two quantities computable by localization. The explicit localization results of these quantities are presented in Section \ref{Sec:localization}. We apply the flat-space limit formula and the localization constraints first to the supergravity solution as a warm-up in Section \ref{Sec:sugra2pt} and then to the lowest few orders of stringy corrections in Section \ref{Sec:stringycorrectionslowest}. We conclude in Section \ref{Sec:outlook} with a brief outlook for future directions. The paper also includes several appendices where we collect some technical details.

\section{Taking the flat-space limit}
\subsection{Appearance of Borel: defect-free case revisited}\label{Sec:defectfreeflatspace}
Let us start by reviewing the defect-free case. The Mellin space representation for an $n$-point scalar correlator in a CFT in $\mathbb{R}^d$ is defined as \cite{Mack:2009mi,Penedones:2010ue}
\begin{equation}
\langle \mathcal{O}_1(x_1)\ldots \mathcal{O}_n(x_n)\rangle=\mathcal{N}_{\rm defect-free} \int [d\delta_{ij}] \mathcal{M}(\delta_{ij})\prod_{i<j}\Gamma(\delta_{ij})x_{ij}^{-2\delta_{ij}}\;,
\end{equation}
where 
\begin{equation}\label{Ndefectfree}
\mathcal{N}_{\rm defect-free}=\frac{\pi^{d/2}}{2}\Gamma\left(\frac{\sum_{i=1}^n\Delta_i-d}{2}\right)\prod_{i=1}^n\frac{1}{\Gamma(\Delta_i)}\;,
\end{equation}
and the variables $\delta_{ij}$ satisfying 
\begin{equation}
\delta_{ij}=\delta_{ji}\;,\quad \delta_{ii}=-\Delta_i\;,\quad \sum_{j}\delta_{ij}=0\;,
\end{equation}
can be viewed as the Mandelstam variables in a fictitious $d+1$ dimensional flat space. The scalar bulk-to-boundary propagator in an $AdS_{d+1}$ space with radius $R$ is 
\begin{equation}\label{GBp}
G_{B\partial}(Z,P)^\Delta=\frac{1}{R^{(d-1)/2}(-2P\cdot Z/R)^\Delta}\;,
\end{equation}
where we have used embedding space coordinates
\begin{equation}
\begin{split}
P={}&(P^+,P^-,\vec{P})=(1,x^2,\vec{x})\;,\\
Z={}&(Z^+,Z^-,\vec{Z})=\frac{1}{z_0}(1,z_0^2+\vec{z}^2,\vec{z})\;.
\end{split}
\end{equation}
To relate to the flat-space amplitude, a formula was given in \cite{Penedones:2010ue}
\begin{equation}\label{flatspacePenedones}
\mathcal{M}(s_{ij})\approx \frac{R^{n(1-d)/2+d+1}}{\Gamma(\frac{1}{2}\sum_i\Delta_i-\frac{d}{2})}\int_0^\infty d\beta \beta^{\frac{1}{2}\sum_i\Delta_i-\frac{d}{2}-1}e^{-\beta}\mathcal{A}\left(S_{ij}=\frac{2\beta}{R^2}s_{ij}\right)\;,\quad R\to\infty\;,
\end{equation}
where
\begin{equation}
\delta_{ij}=\frac{\Delta_i+\Delta_j-s_{ij}}{2}\;,
\end{equation}
and $\mathcal{A}(S_{ij})$ is the flat-space amplitude with Mandelstam variables $S_{ij}$. Note that the flat space  Mandelstam variables $S_{ij}$ are kept finite as we take the flat space limit, $R \to \infty$, so that this formula probes the high energy, or large $s_{ij}$, limit of the Mellin amplitude. Note that the most nontrivial part of this formula is the Borel transform. But it naturally arises in the scalar exchange Witten diagram as was shown in \cite{Penedones:2010ue}. In the following, we review this derivation. 

\begin{figure}[htbp]
\begin{center}
\includegraphics[width=0.62\textwidth]{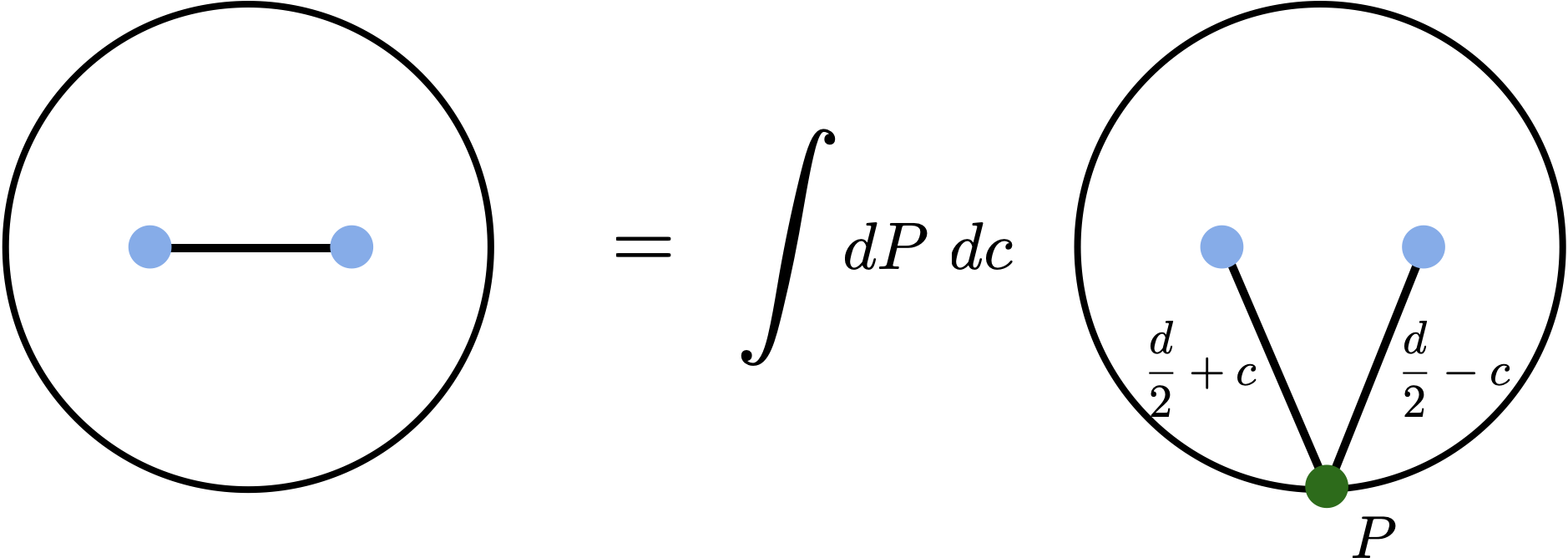}
\caption{Split representation of the bulk-to-bulk propagator. The bulk-to-bulk propagator in $AdS_{d+1}$ is represented as a product of two bulk-to-boundary propagators with dimensions $\frac{d}{2}+c$ and $\frac{d}{2}-c$, and then integrated over the boundary point $P$ in $\mathbb{R}^d$ and the spectral parameter $c$ along the imaginary axis.}
\label{Fig_split}
\end{center}
\end{figure}

We focus on the four-point case for simplicity where the exchanged field has dimension $\Delta$. However, the generalization to an exchange process involving $n$ external points is straightforward. Using the split representation of the bulk-to-bulk propagator, the exchange Witten diagram can be written as the following spectral integral (see Appendix C of \cite{Penedones:2010ue} for details)
\begin{equation}\label{flatspacedefectfree}
\mathcal{M}(s_{ij})=\frac{R^{5-d}}{\Gamma(\frac{\sum_i\Delta_i-d}{2})\Gamma(\frac{\Delta_1+\Delta_2-s}{2})\Gamma(\frac{\Delta_3+\Delta_4-s}{2})}\int_{-i\infty}^{i\infty}\frac{dc}{2\pi i}\frac{l(c)l(-c)}{(\Delta-\frac{d}{2})^2-c^2}\;,
\end{equation}
where $\delta_{12}=\frac{\Delta_1+\Delta_2-s}{2}$, $\delta_{34}=\frac{\Delta_3+\Delta_4-s}{2}$ and
\begin{equation}
l(c)=\frac{\Gamma(\frac{d/2+c-s}{2})\Gamma(\frac{\Delta_1+\Delta_2+c-d/2}{2})\Gamma(\frac{\Delta_3+\Delta_4+c-d/2}{2})}{2\Gamma(c)}\;.
\end{equation}
We are interested in the limit of $R\to\infty$ with squared mass $\Delta(\Delta-d)/R^2$ fixed. This requires $s$ to be large and of the same order as $\Delta^2$. To take this limit, we write $c=iKR$ and require that $K$, $s/R^2$ and $\Delta^2/R^2$ are fixed as $R\to\infty$. Moreover, we will take $s<0$ to avoid the poles	 on the positive real axis. It is straightforward to find
\begin{equation}
\frac{l(iKR)l(-iKR)}{\Gamma(\frac{\Delta_1+\Delta_2-s}{2})\Gamma(\frac{\Delta_3+\Delta_4-s}{2})}\approx \frac{2\pi}{|K|R}\left(-\frac{K^2R^2}{2s}\right)^{\frac{\sum_i\Delta_i-d}{2}}e^{\frac{K^2R^2}{2s}}\;,
\end{equation}
and this gives
\begin{equation}
\mathcal{M}(s)\approx \frac{R^{3-d}}{\Gamma(\frac{\sum_i\Delta_i-d}{2})}\int_0^\infty \frac{dK}{K}\left(-\frac{K^2R^2}{2s}\right)^{\frac{\sum_i\Delta_i-d}{2}}e^{\frac{K^2R^2}{2s}}\frac{1}{\Delta^2/R^2+K^2}\;.
\end{equation}
After making the change of variables $K^2=-2\beta s/R^2$, we get (\ref{flatspacedefectfree}) where $\mathcal{A}$ is the flat-space scalar propagator with squared mass $\Delta^2/R^2$. Although this is a special example, we see that it is nontrivial enough as to capture all the details of the formula (\ref{flatspacedefectfree}).

\subsection{Flat-space limit formula for defect correlators}\label{Sec:flatspaceformuladefect}
We now use the scalar exchange as a trick to obtain a flat-space formula for defect correlators, instead of trying to derive it with the same level of rigor as in the defect-free case. We will first recall some basic kinematics. Here we focus on the simplest observables which are two-point functions of scalar operators inserted away from a defect of general dimension $p$. The generalization to higher-point functions with operators also inserted on the defect is straightforward.  For general $p$, the two-point function depends on two conformal cross ratios\footnote{When $p=d-1$, there is only one cross ratio and the kinematics is the same as a CFT with a boundary or an interface.}  
\begin{equation}\label{defF2ptbosonic}
\llangle \mathcal{O}_{\Delta_1}(x_1)\mathcal{O}_{\Delta_2}(x_2)\rrangle=\frac{\mathcal{F}(\xi,\chi)}{|x_1^i|^{\Delta_1}|x_2^i|^{\Delta_2}}\;,
\end{equation}
where we separate the coordinates into transverse $x^{i=p+1,\ldots d}$ and parallel $x^{a=1,\ldots,p}$ and 
\begin{equation}
\xi=\frac{x_{12}^2}{|x_1^i||x_2^i|}\;,\quad \chi=\frac{2x_1^jx_2^j}{|x_1^i||x_2^i|}\;.
\end{equation}
It follows from the general defect Mellin space formalism \cite{Goncalves:2018fwx} that the two-point function can be written as\footnote{The simpler BCFT Mellin space formalism was developed first in \cite{Rastelli:2017ecj}. It is not a special case of \cite{Goncalves:2018fwx} but the two formalisms are very similar.}
\begin{equation}\label{defMellin}
\mathcal{F}(\xi,\chi)=C_{\Delta_1\Delta_2}\int\frac{d\delta d\gamma}{(2\pi i)^2}\xi^{-\delta}\chi^{-\gamma+\delta}\mathcal{M}(\delta,\gamma)\Gamma(\delta)\Gamma(\gamma-\delta)\prod_{i=1}^2\Gamma\left(\frac{\Delta_i-\gamma}{2}\right)\;,
\end{equation}
where the  Mellin variables $\gamma$, $\delta$ are integrated over the imaginary axes and the normalization $C_{\Delta_1\Delta_2}$ will be fixed later. As was discussed in detail in \cite{Goncalves:2018fwx,Rastelli:2017ecj}, the defect Mellin formalism has many similarities with the defect-free Mellin formalism and can be viewed as the definition of an AdS form factor of particles scattering off an extended object. This is especially evident in theories where the defects are described as $AdS_{p+1}$ probe branes in $AdS_{d+1}$. Some Witten diagrams in this probe brane setup are presented in Fig. \ref{Fig:wittendiagrams}. For example, the Mellin amplitude of a zero-derivative contact Witten diagram (Fig. \ref{Fig_CWD}) is just a constant \cite{Goncalves:2018fwx,Gimenez-Grau:2023fcy}
\begin{equation}
\mathcal{M}_{\rm con,0-der}=\frac{\pi^{\frac{p}{2}}R^{p-d+2}\Gamma(\frac{\Delta_1+\Delta_2-p}{2})}{4\Gamma(\Delta_1)\Gamma(\Delta_2)} C_{\Delta_1\Delta_2}^{-1}\;.
\end{equation}

\begin{figure}
\centering
\begin{subfigure}{0.28\textwidth}
\centering
    \includegraphics[width=0.6\textwidth]{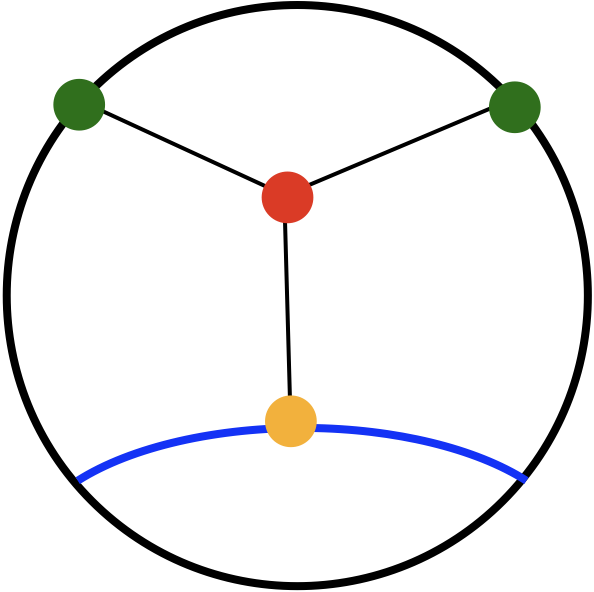}
    \caption{Bulk channel exchange Witten diagrams}
    \label{Fig_bulkEWD}
\end{subfigure}
\hfill
\begin{subfigure}{0.28\textwidth}
\centering
    \includegraphics[width=0.6\textwidth]{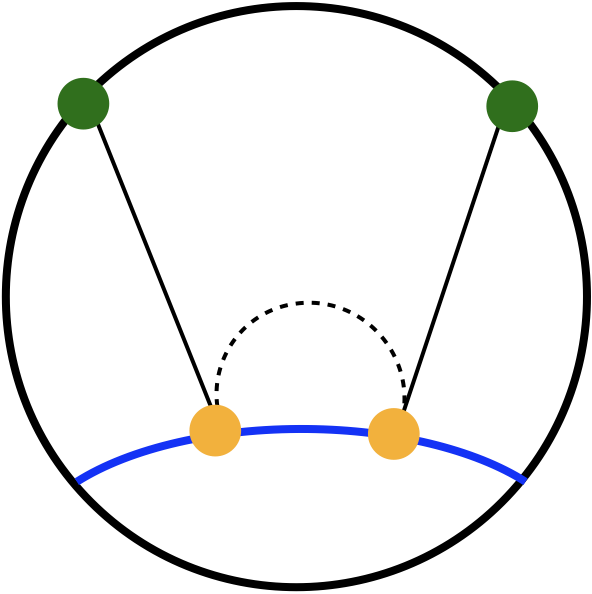}
    \caption{Defect channel exchange Witten diagrams}
    \label{Fig_defectEWD}
\end{subfigure}
\hfill
\begin{subfigure}{0.28\textwidth}
\centering
    \includegraphics[width=0.6\textwidth]{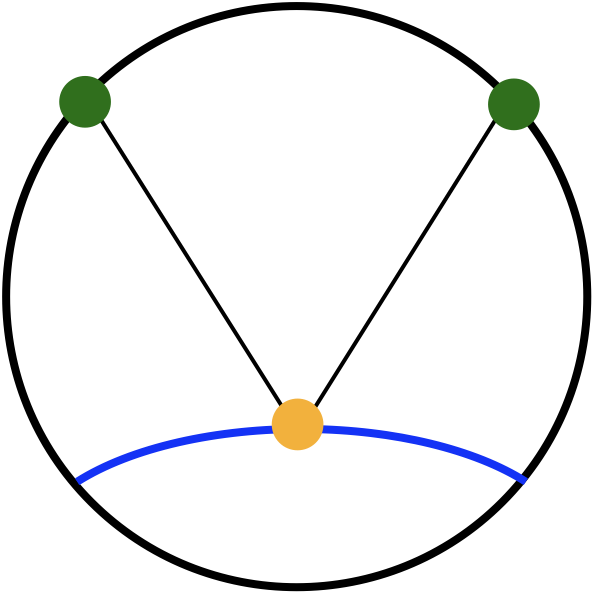}
    \caption{Contact Witten diagrams}
    \label{Fig_CWD}
\end{subfigure}      
\caption{Three types of tree-level Witten diagrams. Here the blue line represents the $AdS_{p+1}$ defect. Solid lines denote propagators in $AdS_{d+1}$ and the dashed line is a propagator in $AdS_{p+1}$.}
\label{Fig:wittendiagrams}
\end{figure}

Our strategy for obtaining the flat-space formula is to use the split representation to study the large $R$ limit of scalar exchange Witten diagrams in Mellin space. A key point here is that we now have two inequivalent types of exchange Witten diagrams: We can exchange the scalar field either in the bulk channel (Fig. \ref{Fig_bulkEWD}) or in the defect channel (Fig. \ref{Fig_defectEWD}). But in the end there should only be one flat-space formula and that we get the same formula from both calculations is a nontrivial consistency check. 

We start with the case where we exchange a scalar of dimension $\Delta$ in the bulk channel. Using the split representation for the bulk-to-bulk propagator, we can express the exchange Witten diagram as the product of a defect-free three-point function and a defect one-point function which are fixed by conformal symmetry. This gives the following spectral representation  \cite{Gimenez-Grau:2023fcy}
\begin{equation}\label{Mbulkexch}
\mathcal{M}_{\Delta_1,\Delta_2}^{\Delta,0}(\delta,\gamma)=\frac{\pi^{p/2}R^{p-d+4}C_{\Delta_1\Delta_2}^{-1}}{\Gamma(\Delta_1)\Gamma(\Delta_2)\Gamma(\delta)\Gamma(\delta-\frac{\Delta_1+\Delta_2-d+p}{2})}\int_{-i\infty}^{i\infty}\frac{d\nu}{2\pi i}\frac{l_b(\nu)l_b(-\nu)}{(\Delta-\frac{d}{2})^2-\nu^2}\;,
\end{equation}
where 
\begin{equation}
l_b(\nu)=\frac{\Gamma(\frac{2\nu+d-2p}{4})\Gamma(\frac{2\nu+2\Delta_1+2\Delta_2-d}{4})\Gamma(\delta+\frac{2\nu-2\Delta_1-2\Delta_2+d}{4})}{4\Gamma(\nu)}\;.
\end{equation}
As in the defect-free case, we introduce new variables
\begin{equation}
\nu=i K R\;,\quad \Delta(\Delta-d)=\mu^2R^2\;,\quad \delta=-\frac{1}{2}S R^2\;,
\end{equation}
where $S$ is a flat-space Mandelstam variable constructed from two momenta $\vec{k}_1$, $\vec{k}_2$
\begin{equation}
S=-\vec{k}_1\cdot \vec{k}_2\;.
\end{equation}
We take the $R\to\infty$ limit keeping $K$, $\mu$, $S$ fixed. We also take $S$ to be negative to avoid the poles on the real axis and find 
\begin{equation}
\frac{l_b(iKR)l_b(-iKR)}{\Gamma(\delta)\Gamma(\delta-\frac{\Delta_1+\Delta_2-d+p}{2})}\approx \frac{\pi}{2KR}\left(-\frac{2S}{K^2}\right)^{\frac{p-\Delta_1-\Delta_2}{2}} e^{\frac{K^2}{2S}}\;.
\end{equation}
Therefore, we can write the Mellin amplitude as
\begin{equation}
\mathcal{M}_{\Delta_1,\Delta_2}^{\Delta,0}\approx \frac{\pi^{p/2}R^{p-d+2}C_{\Delta_1\Delta_2}^{-1}}{2\Gamma(\Delta_1)\Gamma(\Delta_2)}\int_0^\infty \frac{dK}{K} \left(-\frac{2S}{K^2}\right)^{\frac{p-\Delta_1-\Delta_2}{2}} e^{\frac{K^2}{2S}} \frac{1}{K^2+\mu^2}\;,
\end{equation} 
where we have used the symmetry of the integral under $K\leftrightarrow -K$. After the change of variable $K^2\to 4\delta\beta/R^2 $ and recalling $S=-2\delta/R^2$, we get 
\begin{equation}
\mathcal{M}_{\Delta_1,\Delta_2}^{\Delta,0}\approx \frac{\pi^{p/2}R^{p-d+2}C_{\Delta_1\Delta_2}^{-1}}{4\Gamma(\Delta_1)\Gamma(\Delta_2)}\int_0^\infty d\beta \beta^{\frac{1}{2}(\Delta_1+\Delta_2)-\frac{p}{2}-1} e^{-\beta}\frac{1}{\mu^2+4\delta\beta/R^2}\;.
\end{equation}
This is a Borel transformation and we recognize that the transformed function is just the flat-space scalar propagator in the bulk channel $1/(\mu^2-2S)$ as expected. 
 
Let us now look at the defect channel where we exchange a scalar in $AdS_{p+1}$ with dimension $\widehat{\Delta}$. The split representation expresses the defect channel exchange Witten diagram as a product of bulk-defect two functions which are fixed by conformal symmetry. The spectral representation reads \cite{Gimenez-Grau:2023fcy}
\begin{equation}\label{Mdefectexch}
\mathcal{M}_{\Delta_1,\Delta_2}^{\widehat{\Delta},0}(\delta,\gamma)=\frac{\pi^{p/2}R^{p-d+4}C_{\Delta_1\Delta_2}^{-1}}{\Gamma(\Delta_1)\Gamma(\Delta_2)\Gamma(\frac{\Delta_1-\gamma}{2})\Gamma(\frac{\Delta_2-\gamma}{2})}\int_{-i\infty}^{i\infty}\frac{d\nu}{2\pi i}\frac{l_d(\nu)l_d(-\nu)}{(\widehat{\Delta}-\frac{p}{2})^2-\nu^2}\;,
\end{equation}
where 
\begin{equation}
l_d(\nu)=\frac{\Gamma(\frac{2\nu+2\Delta_1-p}{4})\Gamma(\frac{2\nu+2\Delta_2-p}{4})\Gamma(\frac{2\nu+p-2\gamma}{4})}{4\Gamma(\nu)}\;.
\end{equation}
We also make a change of variables 
\begin{equation}
\nu=i K R\;,\quad \widehat{\Delta}(\widehat{\Delta}-p)=\widehat{\mu}^2R^2\;,\quad \gamma=Q R^2\;,
\end{equation}
where $Q$ is the other independent Mandelstam variable
\begin{equation}
Q=-(\vec{k}_{1,\parallel})^2=-(\vec{k}_{2,\parallel})^2\;,
\end{equation}
made of the conserved momenta along the defect. Repeating the same analysis of taking $R$ large with $K$, $\widehat{\mu}$, $Q$ fixed, we find 
\begin{equation}\label{defectmassivescalar}
\mathcal{M}_{\Delta_1,\Delta_2}^{\widehat{\Delta},0}\approx \frac{\pi^{p/2}R^{p-d+2}C_{\Delta_1\Delta_2}^{-1}}{4\Gamma(\Delta_1)\Gamma(\Delta_2)}\int_0^\infty d\beta \beta^{\frac{1}{2}(\Delta_1+\Delta_2)-\frac{p}{2}-1} e^{-\beta}\frac{1}{\widehat{\mu}^2-2\gamma\beta/R^2}\;.
\end{equation} 
This is the same as in the bulk channel case, except that now in flat space we have a scalar propagator in the defect channel $1/(\widehat{\mu}^2-Q)$. 

From these two examples, we conjecture that the flat-space limit formula for a generic defect Mellin amplitude is 
\begin{equation}
\mathcal{M}(\delta,\gamma)\approx \frac{\pi^{p/2}R^{p-d+2}C_{\Delta_1\Delta_2}^{-1}}{4\Gamma(\Delta_1)\Gamma(\Delta_2)} \int_0^\infty d\beta \beta^{\frac{1}{2}(\Delta_1+\Delta_2)-\frac{p}{2}-1}e^{-\beta} \mathcal{A}\left(S=-\frac{2\delta\beta}{R^2},Q=\frac{2\gamma\beta}{R^2}\right)\;,
\end{equation}
with $R$ taken to infinity. We further choose the normalization to be 
\begin{equation}
C_{\Delta_1\Delta_2}=\frac{\pi^{p/2}\Gamma(\frac{\Delta_1+\Delta_2-p}{2})}{4\Gamma(\Delta_1)\Gamma(\Delta_2)}\;.
\end{equation}
so that both the Mellin amplitude and the flat-space amplitude are normalized to one
\begin{equation}
\mathcal{M}_{\rm con,0-der}=R^{p-d+2}\;,\quad \mathcal{A}_{\rm con,0-der}=1\;.
\end{equation}
The formula then becomes
\begin{equation}\label{fslformula}
\mathcal{M}(\delta,\gamma)\approx \frac{R^{p-d+2}}{\Gamma(\frac{\Delta_1+\Delta_2-p}{2})} \int_0^\infty d\beta \beta^{\frac{1}{2}(\Delta_1+\Delta_2)-\frac{p}{2}-1}e^{-\beta} \mathcal{A}\left(S=-\frac{2\delta\beta}{R^2},Q=\frac{2\gamma\beta}{R^2}\right)\;,\; R\to\infty\;.
\end{equation}
Note that the generalization of this formula to higher points is straightforward with the obvious modifications. The generalized formula will have more $\Delta_i$, as well as more Mellin and Mandelstam variables. 

\subsection{Checks}\label{Sec:checks}
Let us now check our flat-space limit formula (\ref{fslformula}) with a few nontrivial examples. 

\subsubsection{Higher-derivative contact Witten diagrams}
We first consider the case of contact Witten diagrams with contracted derivatives in the vertices. Recall that the zero-derivative contact Witten diagram, coming from the vertex $\phi_1\phi_2$, is defined by the integral 
\begin{equation}
W^{\rm con,0-der}_{\Delta_1\Delta_2}=\int \frac{dz_0d^pz}{dz_0^{p+1}}G_{B\partial}^{\Delta_1}(x_1,\hat{z})G_{B\partial}^{\Delta_2}(x_2,\hat{z})\;.
\end{equation}
When the vertex has a pair of contracted derivatives $\nabla_\perp\phi_1\nabla_\perp\phi_2$, where the derivatives are transverse to $AdS_{p+1}$, we get the following two-derivative contact Witten diagram
\begin{equation}
W^{\rm con,2-der}_{\Delta_1\Delta_2}=\int \frac{dz_0d^pz}{dz_0^{p+1}} \frac{z_0^2}{R^2}\partial_i G_{B\partial}^{\Delta_1}(x_1,\hat{z}) \partial_i G_{B\partial}^{\Delta_2}(x_2,\hat{z})\;.
\end{equation}
Using the explicit form of the bulk-to-boundary propagator (\ref{GBp}), it is not difficult to verify the following relation
\begin{equation}\label{Wcon2der}
W^{\rm con,2-der}_{\Delta_1\Delta_2}=\frac{4\Delta_1\Delta_2}{R^2} (x_1^ix_2^i)W^{\rm con,0-der}_{\Delta_1+1\Delta_2+1}\;.
\end{equation}
In Mellin space, this gives the following Mellin amplitude
\begin{equation}
\mathcal{M}^{\rm con,2-der}(\delta,\gamma)=\frac{1}{R^2}(\Delta_1-\Delta_2-p)(\gamma-\delta)\mathcal{M}^{\rm con,0-der}(\delta,\gamma)\;,
\end{equation}
which is linear in the Mellin variables. We insert it into (\ref{fslformula}) and find the flat-space amplitude is
\begin{equation}
\mathcal{A}^{\rm con,2-der}=S+Q\;.
\end{equation}
This is also what we get from Feynman diagrams
\begin{equation}
\mathcal{A}^{\rm con,2-der}=-\vec{k}_{1,\perp}\cdot\vec{k}_{2,\perp}=-\vec{k}_1\cdot \vec{k}_2+\vec{k}_{1,\parallel}\cdot\vec{k}_{2,\parallel}=S+Q\;.
\end{equation}

\subsubsection{Exchange Witten diagrams of spinning fields}
We can also consider exchange Witten diagrams where the exchanged fields have spins. For simplicity, we look at the case where the exchange is in the defect channel. In this case, spins refer to transverse spins, i.e., nontrivial representations charged under the orthogonal $SO(d-p)$ rotations. We can consider a spin-$s$ field $\hat{\phi}^{i_1\ldots i_s}$ coupled to a bulk scalar $\phi$, via the vertex $\hat{\phi}^{i_1\ldots i_s}\partial_{i_1}\ldots\partial_{i_s}\phi$. It leads to a spin-$s$ exchange Witten diagram defined by
\begin{equation}
\widehat{W}^{\widehat{\Delta},s}_{\Delta_1,\Delta_2}=\int \frac{dz_0d^pz}{dz_0^{p+1}}\frac{dw_0d^pw}{dw_0^{p+1}} \frac{(z_0w_0)^s}{R^{2s}}\partial_i^s G_{B\partial}^{\Delta_1}(x_1,\hat{z}) G_{BB}^{\widehat{\Delta}}(z,w) \partial_i^s G_{B\partial}^{\Delta_2}(x_2,\hat{w})\;.
\end{equation}
It is possible to verify using (\ref{GBp}) that this spinning Witten diagram reduces to the scalar one
\begin{equation}
\widehat{W}^{\widehat{\Delta},s}_{\Delta_1,\Delta_2}=(\Delta_1)_s(\Delta_2)_s\left(\frac{4(x_1^ix_2^i)}{R^2}\right)^{s}\widehat{W}^{\widehat{\Delta},0}_{\Delta_1+s,\Delta_2+s}\;.
\end{equation}
The corresponding Mellin amplitude is 
\begin{equation}
\begin{split}
\mathcal{M}^{\widehat{\Delta},s}_{\Delta_1,\Delta_2}(\delta,\gamma)={}&\frac{(\Delta_1)_s(\Delta_2)_s2^s}{R^{2s}}(\gamma-\delta)_s\mathcal{M}^{\widehat{\Delta},0}_{\Delta_1+s,\Delta_2+s}(\delta,\gamma+s)C_{\Delta_1\Delta_2}^{-1}C_{\Delta_1+s,\Delta_2+s}\\
={}&\frac{2^s}{R^{2s}}\left(\frac{\Delta_1+\Delta_2-p}{2}\right)_s(\gamma-\delta)_s\mathcal{M}^{\widehat{\Delta},0}_{\Delta_1+s,\Delta_2+s}(\delta,\gamma+s)\;.
\end{split}
\end{equation}
The flat-space limit formula (\ref{fslformula}) gives in the large $R$ limit
\begin{equation}\label{AsvsA0}
\mathcal{A}^s_{\rm defect}(S,Q)= (Q+S)^s \mathcal{A}^0_{\rm defect}(S,Q)\;,
\end{equation}
which is precisely the relation in flat space. 

Note that this relation is independent of the dimension $\Delta$ of the internal field. In (\ref{defectmassivescalar}), we considered the flat-space limit of defect channel exchange Witten diagrams where the internal line is massive in the flat-space limit. Here let us also  consider the case of exchanging massless particles, i.e., $\widehat{\Delta}$ is fixed and does not scale with $R$. Then by (\ref{AsvsA0}), we will prove the flat-space limit formula for exchanging massless spinning fields in the defect channel. The defect channel scalar exchange Mellin amplitude is given by \cite{Gimenez-Grau:2023fcy}
\begin{equation}
\mathcal{M}^{\widehat{\Delta},0}_{\Delta_1,\Delta_2}=\sum_{n=0}^\infty\frac{V_n}{\gamma-\widehat{\Delta}-2n}\;,
\end{equation}
where 
\begin{equation}
V_n=-R^{p-d+4}\frac{\Gamma(\frac{\widehat{\Delta}+\Delta_1-p}{2})\Gamma(\frac{\widehat{\Delta}+\Delta_2-p}{2})(1+\frac{\widehat{\Delta}-\Delta_1}{2})_n(1+\frac{\widehat{\Delta}-\Delta_2}{2})_n}{2n!\Gamma(\frac{\Delta_1+\Delta_2-p}{2})\Gamma(1+n-\frac{p}{2}+\widehat{\Delta})}\;.
\end{equation}
In the large $R$ limit for fixed $\Delta$ and large $\gamma$, the saddle point of the sum is localized at $\gamma=0$. Therefore, we only need to sum over the numerator and get
\begin{equation}
\mathcal{M}^{\widehat{\Delta},0}_{\Delta_1,\Delta_2}\approx\frac{R^{p-d+4}}{p+2-\Delta_1-\Delta_2}\frac{1}{\gamma}\;.
\end{equation}
This is precisely what we get from (\ref{fslformula}) when $\widehat{\mu}=0$.

\subsubsection{One-loop amplitude}
We can also perform a more sophisticated check at one-loop level. Here we will look at the one-loop example computed in \cite{Chen:2024orp} for the super graviton two-point function in the 6d $\mathcal{N}=(2,0)$ theory in the presence of a $\frac{1}{2}$-BPS surface defect.  The bulk theory is eleven dimensional supergravity in $AdS_7\times S^4$ and the surface defect is dual to an $AdS_3$ subspace with localized degrees of freedom. In this example, one can define a reduced Mellin amplitude which takes into account superconformal symmetry
\begin{equation}\label{M1loop}
\widetilde{\mathcal{M}}_{\rm 1-loop}=\sum_{m,n=0}^\infty \frac{c_{mn}}{(\delta+n)(\gamma-6-2m)}\;.
\end{equation}
Here $c_{mn}$ are numerical coefficients which can be explicitly found in \cite{Chen:2024orp} in terms of ${}_3F_2$ functions.  Because the reduced Mellin amplitude has already taken care of supersymmetry, we expect the particles to become effectively scalars and (\ref{M1loop}) is essentially the Witten diagram in Figure \ref{Fig_1loopAdS} where the propagators are in eleven dimensions. More precisely, in the flat-space limit we expect the one-loop Mellin amplitude to reproduce  the following Feynman integral (Figure \ref{Fig_1loopFeyn})
\begin{equation}\label{Iflat}
I_{\rm flat} = \int d^d \ell_{\perp} \frac{1}{\ell_\perp^2 ((\ell_\perp+k_{1,\perp})^2+k_{1,\|}^2) (\ell_\perp+k_{1,\perp}+k_{2,\perp})^2}\;.
\end{equation}
Here we are interested in the case where $d=8$ for the transverse components and the components parallel to the defects are three dimensional. Moreover, the particles in flat space should be massless. Because momentum is conserved along the defect, we have
\begin{equation}
\ell_\parallel=0\;,\quad k_{1,\parallel}+k_{2,\parallel}=0\;.
\end{equation}
In the following, we will prove their equivalence as a function of the Mandelstam variables but we will not keep track of the overall constants. 

\begin{figure}[htbp]
\begin{center}
\includegraphics[width=0.22\textwidth]{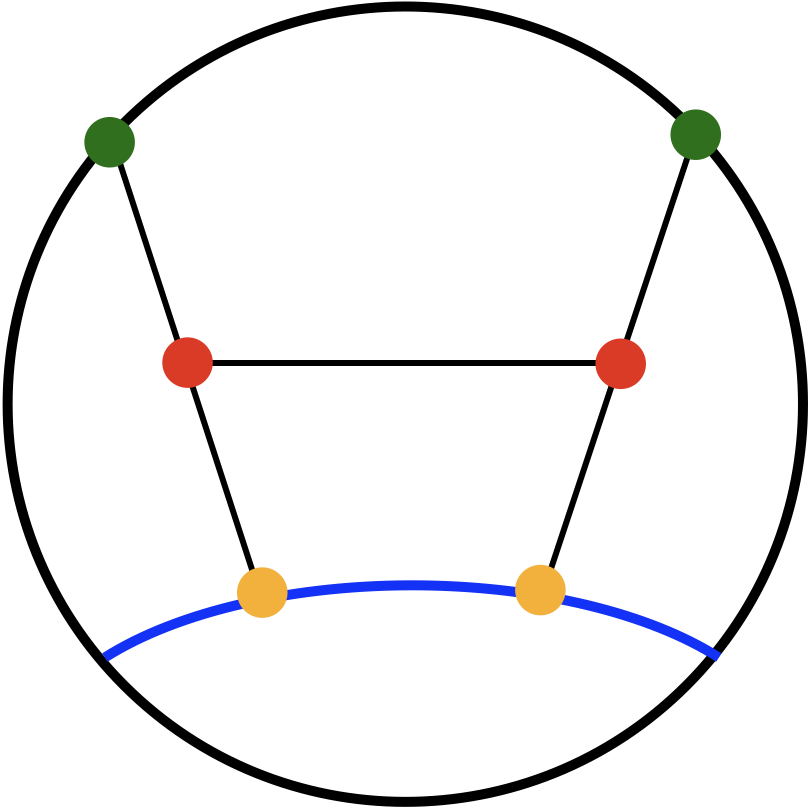}
\caption{A one-loop Witten diagram for a two-point function with a defect.}
\label{Fig_1loopAdS}
\end{center}
\end{figure}

\begin{figure}[htbp]
\begin{center}
\includegraphics[width=0.42\textwidth]{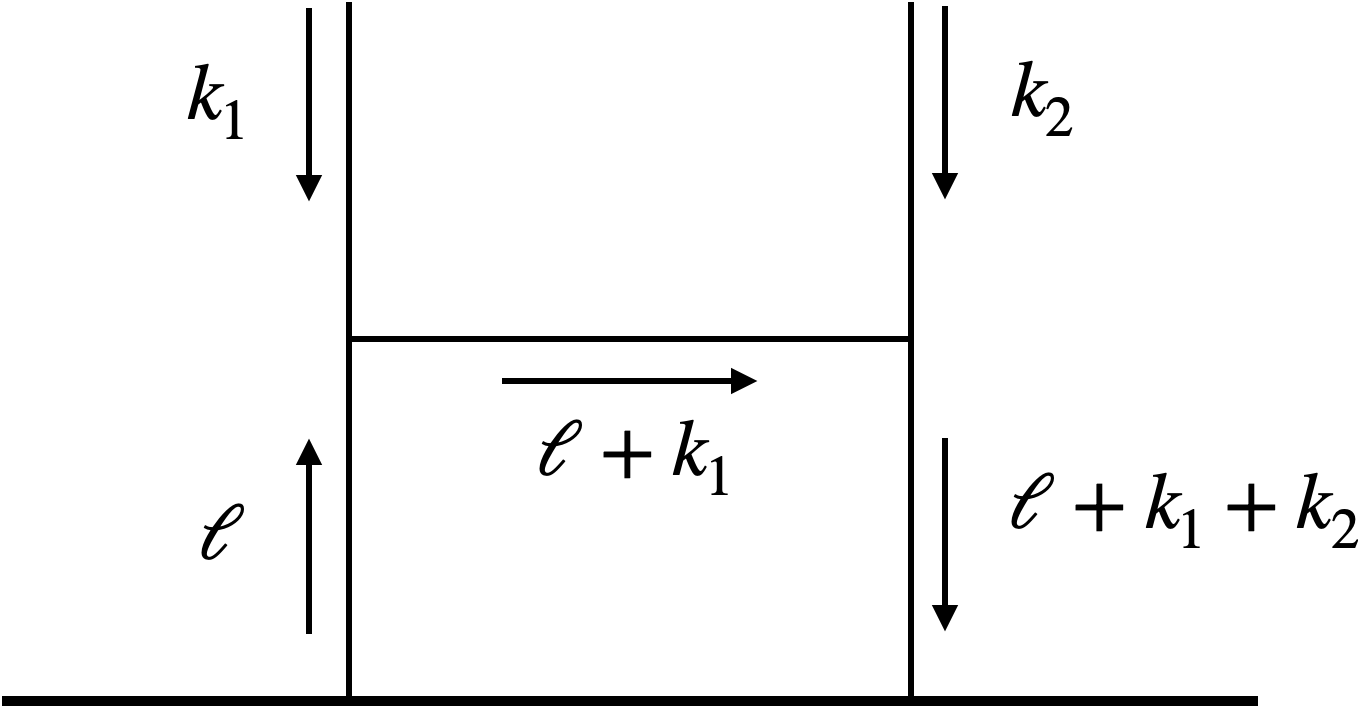}
\caption{A one-loop Feynman diagram in flat space for two particles scattering with a defect.}
\label{Fig_1loopFeyn}
\end{center}
\end{figure}
The dominant contribution as we take $\delta$, $\gamma$ large in (\ref{M1loop}) arises from the region where $m$, $n$ are also large. Finding the large $m$, $n$ behavior of $c_{mn}$ is quite involved and is done in Appendix \ref{App:1loopcheck}. The final result, however, is quite simple. We find that $c_{mn}$ in this limit is
\begin{equation}
c_{mn}\approx \frac{15m^2n^{\frac{3}{2}}}{4(4m+n)^{\frac{5}{2}}}\;.
\end{equation}
Therefore, for large $\delta$, $\gamma$ (\ref{M1loop}) is well approximated by the integral
\begin{equation}
I_{\rm CFT}(\delta,\gamma)=\int_0^\infty dx dy \frac{15x^2y^{\frac{3}{2}}}{4(4x+y)^{\frac{5}{2}}}\frac{1}{(\delta+y)(\gamma-2x)}\;,
\end{equation}
where we will also assume $\delta>0$ and $\gamma<0$ to avoid the branch cuts. This integral is clearly divergent and later we will consider its derivatives to regularize it. We can first use the Schwinger parameterization and then integrate over $x$ and $y$. We get 
 \begin{equation}
 \begin{split}
I_{\rm CFT}(\delta,\gamma)={}&\int_0^\infty dt_1dt_2 e^{\gamma t_2-\delta t_1}\frac{5}{32\sqrt{t_2}(2t_1-t_2)^{\frac{9}{2}}}\bigg(-\sqrt{2t_1-t_2}(25t_2^{\frac{3}{2}}+55t_1t_2^{\frac{1}{2}})\\
{}&\quad\quad\quad\quad+6(3t_1^2+12t_1t_2+2t_2^2){\rm tan}^{-1}\bigg(\sqrt{\frac{2t_1}{t_2}-1}\bigg)\bigg)\;.
\end{split}
\end{equation}
We can then do a change of variables into $t_1=\beta t$, $t_2=(1-\beta)t$ with $\beta\in[0,1]$. The integral over $t$ is simple to perform but it becomes convergent only after taking two derivatives. One can show the integral satisfies the following identity
\begin{equation}\label{diffICFT1}
(\delta\partial_\delta+\gamma\partial_\gamma)(\kappa_1\partial_\delta+\kappa_2\partial_\gamma)I_{\rm CFT}(\delta,\gamma)=\frac{5(\kappa_1-\kappa_2)}{32}\;,
\end{equation}
where $\kappa_1$, $\kappa_2$ are arbitrary constants. One can also check that another independent combination of derivatives gives 
\begin{equation}\label{diffICFT2}
(\partial_\gamma^2+4\partial_\gamma\partial_\delta+4\partial_\delta^2)I_{\rm CFT}(\delta,\gamma)=\frac{1}{\delta}F(\rho)\;,
\end{equation}
where
\begin{equation}
\rho=\frac{\gamma}{\gamma-\delta}\;,
\end{equation}
and
\begin{equation}
\begin{split}
F(\rho)={}&\frac{5 \sqrt{1-\rho }}{64 (\rho +1)^{5/2}}\bigg(10 (16 \rho -5) \sqrt{1-\rho ^2} \log \bigg(\frac{2}{\rho }-2\bigg)+70 \sqrt{1-\rho } (\rho +1)^{3/2}\\
{}&+(\rho  (17 \rho -16)+2) \bigg(12 \text{Li}_2\bigg(\frac{1-\sqrt{1-\rho ^2}}{\rho }\bigg)+3 \log ^2\bigg(\frac{1-\sqrt{1-\rho ^2}}{\rho }\bigg)+4 \pi ^2\bigg)\bigg)\;.
\end{split}
\end{equation}

For the flat-space integral (\ref{Iflat}), we can also use the Schwinger parameterization 
\begin{equation}
I_{\rm flat}(\delta,\gamma)=\int_0^\infty d\alpha_1d\alpha_2d\alpha_3\int d^d\ell_\perp e^{-\alpha_1\ell_\perp^2-\alpha_2((\ell_\perp+k_{1,\perp})^2+k_{1,\parallel}^2)-\alpha_3(\ell_\perp+k_{1,\perp}+k_{2,\perp})^2}\;.
\end{equation}
It is easy to perform the Gaussian integral with respect to $\ell_\perp$ and get
\begin{equation}
I_{\rm flat}(\delta,\gamma)=\int_0^\infty dt\int_0^1d\beta_1d\beta_2d\beta_3\delta(\sum_i\beta_i-1)t^{2-\frac{d}{2}}e^{t(-2\beta_1\beta_3\delta+\beta_2^2\gamma)}\;,
\end{equation}
where we have made the change of variables $\alpha_i=\beta_i t$ with $t=\alpha_1+\alpha_2+\alpha_3$. For $d=8$, the integral over $t$ is again divergent but can be regularized by taking two derivatives. Again, one can explicitly check that 
\begin{equation}
(\delta\partial_\delta+\gamma\partial_\gamma)(\kappa_1\partial_\delta+\kappa_2\partial_\gamma)I_{\rm flat}(\delta,\gamma)=\frac{\kappa_1-\kappa_2}{12}\;,
\end{equation}
\begin{equation}
(\partial_\gamma^2+4\partial_\gamma\partial_\delta+4\partial_\delta^2)I_{\rm CFT}(\delta,\gamma)=\frac{8}{15\delta}F(\rho)\;,
\end{equation}
which are exactly (\ref{diffICFT1}) and (\ref{diffICFT2}) up to an overall constant! This proves that $I_{\rm CFT}(\delta,\gamma)$ and $I_{\rm flat}(\delta,\gamma)$ are essentially the same up to divergent ambiguities which are degree 1 polynomials in $\delta$ and $\gamma$. 

\section{Stringy corrections: gravitons scattering off a D1 brane}\label{Sec:stringycorrections}
\subsection{Twisted and integrated correlators}\label{Sec:twistedandintegrated}
In this section, we consider the two-point function of the stress tensor multiplet in 4d $\mathcal{N}=4$ SYM when a half-BPS line defect is present. These operators are defined as 
\begin{equation}
S(x,u)=u^m u^n N_S\left[\phi^I_m(x)\phi^I_n(x)-\frac{1}{6}\delta_{mn}\phi^I_p(x)\phi^I_p(x)\right]\;,
\end{equation}
where $u$ is a null polarization vector of the $SO(6)$ R-symmetry group. The adjoint index $I$ of the gauge group $SU(N)$ takes value $I=1,\ldots,N^2-1$ and the normalization factor $N_S$ is taken to be 
\begin{equation}
N_S=\frac{2\sqrt{2}\pi^2}{g_{YM}^2\sqrt{N^2-1}}\;,
\end{equation}
such that the two-point function without the defect is normalized to one
\begin{equation}
\langle S(x_1,u_1)S(x_2,u_2)\rangle = \frac{(u_1\cdot u_2)^2}{x_{12}^4}\;.
\end{equation} 
The two-point function in the presence of a $\frac{1}{2}$-BPS line defect can be written as a function of three cross ratios
\begin{equation}\label{defF}
\begin{split}
\llangle S(x_1,u_1)S(x_2,u_2)\rrangle ={}&\frac{(u_1\cdot \theta)^2(u_2\cdot \theta)^2}{|x_1^i|^2|x_2^i|^2}\mathcal{F}(\zeta,\eta;\sigma)\\
={}&\frac{(u_1\cdot \theta)^2(u_2\cdot \theta)^2}{|x_1^i|^2|x_2^i|^2}\left(\mathcal{F}_0(\zeta,\eta)+\sigma\mathcal{F}_1(\zeta,\eta)+\sigma^2\mathcal{F}_2(\zeta,\eta)\right)\;,
\end{split}
\end{equation}
where it is convenient to make a change of variables as in \cite{Dempsey:2024vkf} for the conformal cross ratios
\begin{equation}
\zeta=\frac{(\tau_1-\tau_2)^2+|x_1^i|^2+|x_2^i|^2}{2|x_1^i||x_2^i|}=\frac{\xi+\chi}{2}\;,\quad \eta=\frac{x_1^jx_2^j}{|x_1^i||x_2^i|}=\frac{\chi}{2}\;,
\end{equation}
and we have denoted the coordinate parallel to the line defect by $\tau$. An advantage of working with $\zeta$ and $\eta$ is that the allowed value of $(\zeta,\eta)$ in Euclidean signature is the product  $[1,\infty)\times [-1,1]$. Additionally, we also define the R-symmetry cross ratio
\begin{equation}
\sigma=\frac{(u_1\cdot u_2)}{(u_1\cdot\theta)(u_2\cdot\theta)}\;,
\end{equation}
using a fixed unit vector $\theta$ with $\theta\cdot\theta=1$ which captures the breaking of R-symmetry by the line defect from $SO(6)$ to $SO(5)$.

Superconformal symmetry has further consequences for the correlator. The first is the superconformal Ward identities which are constraints imposed by fermionic generators. They take the form \cite{Liendo:2016ymz}
\begin{equation}\label{scfWardid}
\left(\partial_z+\frac{1}{2}\partial_\alpha\right)\mathcal{F}(z,\bar{z};\alpha)\bigg|_{\alpha=z}=0\;,
\end{equation}
and similarly with $z\leftrightarrow\bar{z}$. Here we have made the change of variables
\begin{equation}
\zeta=\frac{1+z\bar{z}}{2\sqrt{z\bar{z}}}\;,\quad\eta=\frac{z+\bar{z}}{2\sqrt{z\bar{z}}}\;,\quad \sigma=-\frac{(1-\alpha)^2}{2\alpha}\;.
\end{equation}
The physical meaning of the $z$, $\bar{z}$ coordinates is as follows. We can use conformal symmetry to move the two operators to be on a 2d plane (spanned by the first two components)
\begin{equation}
x_1=(\frac{z-\bar{z}}{2i},\frac{z+\bar{z}}{2},0,0)\;,\quad x_2=(0,1,0,0)\;,
\end{equation}
and place the line defect at $x^2=x^3=x^4=0$. It is clear from (\ref{scfWardid}) that we will get a number if we set $z=\bar{z}=\alpha$. More precisely, this quantity is almost topological but  depends on the sign of $z$
\begin{equation}
\mathcal{F}_{\rm top}(\vartheta)=\mathcal{F}(z,z;z)\;,
\end{equation}
where $\vartheta={\rm sgn} z$. In other words, $\mathcal{F}_{\rm top}$ can be different depending on whether the two operators are on the same side or different sides of the line defect in this 2d plane. 

The second consequence of superconformal symmetry is that there are certain quantities exactly computable by supersymmetric localization \cite{Pufu:2023vwo,Billo:2023ncz,Dempsey:2024vkf,Billo:2024kri} as functions of the complex coupling $\tau=\tau_1+i\tau_2=\frac{\theta}{2\pi}+\frac{4\pi i}{g^2}$. As was explained in \cite{Dempsey:2024vkf}, there are two independent quantities, obtained by taking derivatives of the expectation value of the supersymmetric line defect in $\mathcal{N}=2^*$ SYM. The first arises from the derivatives with respect to the mass
\begin{equation}\label{defIL}
\mathcal{I}_{\mathbb{L}}(\tau,\bar{\tau})=\partial_m^2\log\langle \mathbb{L}\rangle\big|_{m=0}\;,
\end{equation}
and the second arises from the derivatives with respect to the the coupling
\begin{equation}\label{defEL}
\mathcal{E}_{\mathbb{L}}(\tau,\bar{\tau})=\partial_\tau\partial_{\bar{\tau}}\log\langle \mathbb{L}\rangle\big|_{m=0}\;.
\end{equation}  
The first object is related to the integrated two-point function \cite{Pufu:2023vwo,Billo:2023ncz,Dempsey:2024vkf,Billo:2024kri}. The relevant part is the component $\mathcal{F}_2$ in (\ref{defF}) and the precise relation is \cite{Dempsey:2024vkf}
\begin{equation}\label{integrated2pt}
\mathcal{I}_{\mathbb{L}}=-16\pi^4\int_1^\infty d\zeta \int_{-1}^1 d\eta \frac{N^2}{4\pi^4} \mathcal{F}_2(\zeta,\eta)\;.
\end{equation}
The second object is related to the topologically twisted correlator \cite{Pufu:2023vwo}
\begin{equation}
\mathcal{E}_{\mathbb{L}}(\tau,\bar{\tau})=\frac{c}{8\tau_2^2}\left(\mathcal{F}_{\rm top}(-1)-a_{\mathbb{L}}(\tau,\bar{\tau})^2\right)
\end{equation}
where $c=(N^2-1)/4$ and $a_{\mathbb{L}}(\tau,\bar{\tau})$ is the one-point function coefficient
\begin{equation}
\llangle S(x,u)\rrangle=a_{\mathbb{L}}(\tau,\bar{\tau})\frac{(u\cdot\theta)^2}{|x^i|^2}\;,
\end{equation}
which can also be computed by localization and is given by
\begin{equation}\label{aL}
a_{\mathbb{L}}=-i\frac{\tau_2}{\sqrt{2c}}\partial_\tau \log\langle \mathbb{L}\rangle\big|_{m=0}\;.
\end{equation}
In other words, $\mathcal{E}_{\mathbb{L}}$ computes the connected part of the twisted correlator\footnote{We think that in \cite{Pufu:2023vwo} there is a factor of $-i/4$ missing in $a_{\mathbb{L}}$ and a factor of $16$ missing in $\mathcal{E}_{\mathbb{L}}$.}
\begin{equation}\label{ELconn}
\mathcal{E}_{\mathbb{L}}(\tau,\bar{\tau})=\frac{2c}{\tau_2^2}\mathcal{F}_{\rm top,conn}(-1)\;,
\end{equation}
where we subtract the identity operator exchange in the defect channel, i.e., the product of one-point functions. Note that so far the discussion has been kept very general and depends only on the kinematics of the two-point function. Therefore, they apply to any $(p,q)$ Wilson-'t Hooft line defects. Together with the flat-space limit (\ref{fslformula}), we now have three complementary nontrivial conditions. 

\subsection{Results from localization}\label{Sec:localization}
In this paper, we are particularly interested in the case where we scatter gravitons off a D1 brane in AdS. This case corresponds to the line defect $\mathbb{L}$ being a 't Hooft line. In this subsection, we will summarize the results about the two quantities $\mathcal{I}_{\mathbb{T}}$ and $\mathcal{E}_{\mathbb{T}}$ which we will use later.

For the integrated correlator (\ref{defIL}), the result has already been computed in \cite{Pufu:2023vwo} and reads\footnote{In \cite{Pufu:2023vwo}, the expansion is with respect to the large dual 't Hooft coupling $\tilde{\lambda}$. This is because \cite{Pufu:2023vwo} started with Wilson loops and used the fact that Wilson loops and 't Hooft loops are related by an S-duality. S-duality changes the coupling of the theory from $\tau$ to $\tilde{\tau}=-1/\tau$, or $g\to \tilde{g}=4\pi/g$ (we focus on zero theta angle), and the dual 't Hooft coupling is defined as $\tilde{\lambda}=\tilde{g}^2N$. Here we keep the coupling of the theory to be $\tau$ (or 't Hooft coupling to be $\lambda$). This amounts to starting with the Wilson loop results with coupling $\tilde{\tau}$, and is equivalent to removing the tilde in the results for the 't Hooft loop in \cite{Pufu:2023vwo}.}
\begin{equation}\label{localIT}
\mathcal{I}_{\mathbb{T}}=N\left(\frac{4\pi}{\lambda^{\frac{1}{2}}}-\frac{2\pi^3}{\lambda^{\frac{3}{2}}}+\frac{24\pi\zeta(3)}{\lambda^2}+\frac{\pi^5}{6\lambda^{\frac{5}{2}}}+\mathcal{O}(\lambda^{-3})\right)+\mathcal{O}(N^0)\;.
\end{equation}
For $\mathcal{E}_{\mathbb{T}}$ (\ref{defEL}), we can start with the vacuum expectation value for the 't Hooft loop \cite{Drukker:2000rr}
\begin{equation}
\langle \mathbb{T}\rangle(\tau)=\langle \mathbb{W}\rangle(-1/\tau)=\frac{1}{N}e^{\frac{N-1}{N}\frac{\pi|\tau|^2}{2\tau_2}}L_{N-1}^1\left(-\frac{\pi|\tau|^2}{\tau_2}\right)\;,
\end{equation}
and then take derivatives with respect to the coupling. Here $L_n^k$ is the associated Laguerre polynomial 
\begin{equation}
L_n^k(x)=\frac{e^x}{x^kn!}\frac{d^n}{dx^n}e^{-x}x^{n+k}=(-1)^k\frac{d^k}{dx^k}L_{n+k}(x)\;.
\end{equation}
We get in the large $N$ limit
\begin{equation}\label{localET}
\begin{split}
\mathcal{E}_{\mathbb{T}}={}&\frac{1}{N}\frac{\lambda(3\lambda+4\pi^2)}{64\pi\sqrt{\lambda+\pi^2}}+\mathcal{O}(N^{-2})\\
={}&\frac{1}{N}\left(\frac{3\lambda^{\frac{3}{2}}}{64\pi}+\frac{5\pi \lambda^{\frac{1}{2}}}{128}-\frac{7\pi^3}{512\lambda^{\frac{1}{2}}}+\frac{9\pi^5}{1024\lambda^{\frac{3}{2}}}+\mathcal{O}(\lambda^{-\frac{5}{2}})\right)+\mathcal{O}(N^{-2})\;.
\end{split}
\end{equation}
Taking this limit requires some properties of the Laguerre polynomial which we present in Appendix \ref{App:LaguerrelargeN}.

\subsection{Supergravity approximation}\label{Sec:sugra2pt}
In this subsection, we first consider the leading order contribution to the two-point function at strong coupling captured by supergravity, before moving onto stringy corrections. A minor point to notice is that the flat-space limit formula (\ref{fslformula}) is modified by a multiplicative factor when there is an internal space. 

The defect two-point function in the supergravity limit was first bootstrapped in \cite{Barrat:2021yvp} using a dispersion relation together with an ad hoc ansatz, and later reproduced more systematically in \cite{Gimenez-Grau:2023fcy} by using analytic bootstrap techniques similar to those of \cite{Rastelli:2016nze,Rastelli:2017udc}. Although in \cite{Barrat:2021yvp,Gimenez-Grau:2023fcy} the line defect considered is a Wilson loop (dual to F1 string in AdS), the supergravity ansatz in \cite{Gimenez-Grau:2023fcy} is insensitive to this information. Therefore, the uniqueness of the solution to the superconformal Ward identities dictates that, in the supergravity approximation, the defect two-point function for the 't Hooft loop is the same as for the Wilson loop, up to an overall constant. To write down the supergravity two-point function, it is useful to introduce a new variable
\begin{equation}
\zeta=\frac{1}{2}\left(r+\frac{1}{r}\right)\;,
\end{equation}
where $r$ is restricted to $(0,1]$. The defect correlator reads \cite{Gimenez-Grau:2023fcy}
\begin{equation}
\mathcal{F}_{\rm sugra}=\mathcal{N}_{\rm sugra}^{\mathbb{T}}\left(\sigma\left(1-\frac{\sigma}{6}\right)\mathcal{P}^{2,0}_{22}+\frac{\sigma^2}{360}\mathcal{P}^{4,2}_{22}+(\sigma-1)\widehat{\mathcal{W}}^{1,0}_{22}+\frac{1}{4}\widehat{\mathcal{W}}^{2,1}_{22}+\frac{1-2\sigma}{\pi}\mathcal{W}^{\rm con,0-der}_{22}\right)\;.
\end{equation}
Here $\mathcal{P}^{\Delta,\ell}_{\Delta_1\Delta_2}$ are suitably normalized bulk channel exchange Witten diagrams (with the addition of particular contact Witten diagrams)\footnote{They are the Polyakov-Regge blocks \cite{Gimenez-Grau:2023fcy} in the terminology of \cite{Mazac:2018biw,Mazac:2019shk} and the contact diagrams are added to improve the Regge behavior. For all the Witten diagrams, we have extracted a kinematic factor according to (\ref{defF2ptbosonic}) so that they are functions of the conformal cross ratios.}
\begin{equation}
\begin{split}
\mathcal{P}^{2,0}_{22}={}&\frac{2 r^2 \log r}{(r-1) (r+1) \left(r^2-r \chi +1\right)}\;,\\
\mathcal{P}^{4,2}_{22}={}&\frac{60 r^2 \left(-3 r^4+2 \left(r^4+4 r^2+1\right) \log (r)+3\right)}{\left(r^2-1\right)^3 \left(r^2-r \chi +1\right)}\;.
\end{split}
\end{equation}
and the defect channel exchange Witten diagrams 
 $\widehat{\mathcal{W}}^{\widehat{\Delta},s}_{\Delta_1\Delta_2}$ read
 \begin{equation}
\begin{split}
{}&\widehat{\mathcal{W}}^{1,0}_{22}=\log (r+1)-\frac{r^2 \log r}{r^2-1}\;,\\
{}&\widehat{\mathcal{W}}^{2,1}_{22}=\chi  \left(\frac{-2 r^4+r^3+4 r^2+r-2}{2 \left(r^2-1\right)^2}-\frac{\left(r^4-2 r^2+3\right) r^3 \log r}{\left(r^2-1\right)^3}+\left(r+\frac{1}{r}\right) \log (r+1)\right)\;.
\end{split}
\end{equation}
Finally, a generic contact Witten diagram (for line defects) can be evaluated as 
\begin{equation}\label{Wcon0der}
\mathcal{W}^{\rm con,0-der}_{\Delta_1\Delta_2}=\frac{\pi\Gamma(\frac{\Delta_1+\Delta_2-1}{2})}{2^{\Delta_1+\Delta_2}\Gamma(\frac{\Delta_1+\Delta_2+1}{2})}{}_2F_1\left(\Delta_1,\Delta_2,\frac{\Delta_1+\Delta_2+1}{2},-\frac{\zeta-1}{2}\right)\;,
\end{equation}
which reduces in the case at hand to
\begin{equation}
\mathcal{W}^{\rm con,0-der}_{22}=\frac{\pi  r^2 \left(-r^2+\left(r^2+1\right) \log (r)+1\right)}{2 \left(r^2-1\right)^3}\;.
\end{equation}
It is straightforward to compute the twisted correlator and we find
\begin{equation}
\mathcal{F}_{\rm sugra,top}(-1)=\frac{3}{8}\mathcal{N}_{\rm sugra}^{\mathbb{T}}\;.
\end{equation}
Using (\ref{integrated2pt}), we also get the integrated correlator
\begin{equation}
\mathcal{I}_{\mathbb{T},{\rm sugra}}=N^2 \mathcal{N}_{\rm sugra}^{\mathbb{T}}\;.
\end{equation}
Comparing these results with (\ref{localIT}) and (\ref{localET}), we find 
\begin{equation}
\mathcal{N}_{\rm sugra}^{\mathbb{T}}=\frac{4\pi}{\sqrt{\lambda}N}\;.
\end{equation}

Let us now consider the supergravity two-point function in Mellin space. The Mellin amplitude is
\begin{equation}
\begin{split}
\mathcal{M}_{\rm sugra}={}&\frac{\mathcal{N}_{\rm sugra}^{\mathbb{T}}}{\pi}\bigg(\frac{\sigma  (\sigma  (\gamma -\delta )-\sigma +4)}{\delta -1}-\frac{2 (\gamma -\delta )+4 \sigma -4}{\gamma }+\frac{\gamma -\delta }{\gamma +2}+(\sigma -1)^2\\
{}&+\frac{2^{\gamma +2} \Gamma (-\gamma ) (\frac{\delta +2}{\gamma +2}-\sigma)}{\Gamma (\frac{2-\gamma }{2})^2}\bigg)\;,
\end{split}
\end{equation}
which behaves for large $\delta$, $\gamma$ as
\begin{equation}\label{Msugraflatspace}
\mathcal{M}_{\rm sugra}\approx \frac{\mathcal{N}_{\rm sugra}^{\mathbb{T}}}{\pi} \frac{(\delta-\gamma\sigma)^2}{\gamma\delta}\;.
\end{equation}
Plugging it in the flat-space formula (\ref{fslformula}) and introducing another constant factor $C_{S^5}$ on the RHS to account for the internal space, we find the expected flat-space result is
\begin{equation}\label{AsugraAdS5}
\mathcal{A}_{\rm sugra}^{AdS_5\times S^5}= -RC_{S^5}^{-1} \frac{\mathcal{N}_{\rm sugra}^{\mathbb{T}}}{\pi} \frac{(S+Q\sigma)^2}{QS}\;.
\end{equation}
This should be compared with the string theory calculation in flat space \cite{Klebanov:1995ni,Hashimoto:1996bf,Garousi:1996ad}. The 1-to-1 tree-level amplitude of a graviton scattering off a D-brane can be computed as a worldsheet integral on a disk with two punctures and reads
\begin{equation}
\mathcal{A}^{\rm disk}_{h\to h}(p_1,e_1;p_2,e_2)=-\frac{1}{8}T_{\rm D1}\ell_s^4\frac{\Gamma(\frac{\ell_s^2s}{2})\Gamma(\frac{\ell_s^2t}{2})}{\Gamma(1+\frac{\ell_s^2s+\ell_s^2t}{2})}(sa_1-ta_2)\;,
\end{equation}
where 
\begin{equation}
\begin{split}
a_1={}&{\rm Tr}(e_1\cdot D)p_1\cdot e_2\cdot p_1-p_1\cdot e_2\cdot D\cdot e_1\cdot p_2-2p_1\cdot e_2\cdot e_1 \cdot D \cdot p_1-p_1\cdot e_2 \cdot e_1 \cdot p_2\\
{}&+\frac{s}{2}{\rm Tr}(e_1 \cdot e_2)+(1\leftrightarrow 2)\;,\\
a_2={}&{\rm Tr}(e_1\cdot D)(p_1\cdot e_2\cdot D\cdot p_2+p_2\cdot D\cdot e_2\cdot p_1+p_2\cdot D\cdot e_2\cdot D\cdot p_2)\\
{}&+p_1\cdot D\cdot e_1\cdot D\cdot e_2\cdot D\cdot p_2-p_2\cdot D\cdot e_2\cdot e_1\cdot D\cdot p_1+\frac{s}{2}{\rm Tr}(e_1\cdot D\cdot e_2\cdot D)\\
{}&-\frac{s}{2}{\rm Tr}(e_1\cdot e_2)-\frac{s+t}{2}{\rm Tr}(e_1\cdot D){\rm Tr}(e_2\cdot D)+(1\leftrightarrow 2)\;.
\end{split}
\end{equation}
Here $e_{i,\mu\nu}$ are polarization tensors of gravitons and $D^\mu{}_\nu={\rm diag}(1,1,-1,\ldots,-1)$ when the D1 brane is along the first two directions with plus signs. The Mandelstam variables $s$, $t$ are related to our definitions of $S$, $Q$ by
\begin{equation}
t=-S\;,\quad s=-2Q\;.
\end{equation}
When we obtain this amplitude from AdS by taking the flat-space limit, it is important to note that we are not in the generic case but are restricted to a special kinematic configuration \cite{Chester:2018aca,Alday:2021odx}. The flat-space polarizations are related to the R-symmetry polarization vectors by 
\begin{equation}
e_{i,\mu\nu}=u_{i,\mu}u_{i,\nu}\;,
\end{equation}
where we lift the six dimensional $u_i$ into ten dimensions by adding zeros. The matrix $D$ can be expressed as
\begin{equation}
D_{\mu\nu}=2(\theta_\mu\theta_\nu+\tau_\mu\tau_\nu)-\eta_{\mu\nu}\;,
\end{equation}
where $\theta$ is the fixed vector (also lifted to 10d) we used to break the R-symmetry from $SO(6)$ to $SO(5)$ and $\tau$ is another unit vector orthogonal to $\theta$. Furthermore, we should impose the transversality conditions
\begin{equation}\label{orthcondi}
\theta\cdot p_i=0\;,\quad u_i\cdot p_j=0\;,\quad u_i\cdot \tau=0\;,
\end{equation}
for any $i$ and $j$ because these vectors live in orthogonal spaces coming from the two factors of $AdS_5\times S^5$. This greatly simplifies the coefficients $a_1$ and $a_2$, and the only nonzero terms are 
\begin{equation}
\begin{split}
{}&{\rm Tr}(e_i\cdot D)=2(u_i\cdot \theta)^2\;,\quad {\rm Tr}(e_1\cdot e_2)=(u_1\cdot u_2)^2\;,\\
{}&{\rm Tr}(e_1\cdot D\cdot e_2\cdot D)=(2(u_1\cdot \theta)(u_2\cdot\theta)-(u_1\cdot u_2))^2\;.
\end{split}
\end{equation}
We get
\begin{equation}
a_1=s(u_1\cdot u_2)^2\;,\quad a_2=-4t(u_1\cdot\theta)^2(u_2\cdot\theta)^2-4s(u_1\cdot\theta)(u_2\cdot\theta)(u_1\cdot u_2)\;.
\end{equation}
Therefore, the flat-space amplitude becomes 
\begin{equation}\label{Adiskorth}
\mathcal{A}^{\rm disk}_{h\to h}(p_1,e_1;p_2,e_2)=-\frac{1}{8}T_{\rm D1}\ell_s^4\frac{\Gamma(\frac{\ell_s^2s}{2})\Gamma(\frac{\ell_s^2t}{2})}{\Gamma(1+\frac{\ell_s^2s+\ell_s^2 t}{2})}(s(u_1\cdot u_2)+2t(u_1\cdot \theta)(u_2\cdot\theta))^2\;.
\end{equation}
In the supergravity limit $\ell_s\to0$, this gives precisely (\ref{AsugraAdS5}) and also determines
\begin{equation}
C_{S^5}=\frac{\mathcal{N}_{\rm sugra}^{\mathbb{T}} R}{\pi T_{\rm D1}}\;.
\end{equation}

\subsection{Stringy corrections at lowest orders}\label{Sec:stringycorrectionslowest}

\subsubsection{Superconformal higher-derivative terms}
The stringy corrections to the defect two-point function are manifested in two ways. The first manifestation is in the correction to one-point functions. As was shown in \cite{Gimenez-Grau:2023fcy}, the supergravity two-point function is fixed by superconformal symmetry up to an overall coefficient. We can decompose the correlator into conformal blocks in the bulk channel and take this overall coefficient to be the decomposition coefficient of the super graviton $S$ which is the product of the three-point function coefficient of $\langle SSS\rangle$ without defects and the defect one-point function coefficient $\llangle S\rrangle$.\footnote{Of course, we can also see it by decomposing the two-point function into defect channel conformal blocks. The results will be equivalent but we need to compute the bulk-defect two-point function.} While it is well-known that the three-point function is protected by non-renormalization theorems, the one-point function is not. Therefore, the corrections to the defect two-point function will receive a contribution to all orders in $1/\lambda$ which is proportional to the supergravity two-point function, coming from the one-point function.\footnote{Note this does not happen for defect-free four-point functions because the decomposition coefficients are proportional to the product of three-point function coefficients which are protected for $\frac{1}{2}$-BPS operators.} This part is fixed exactly for all $\lambda$ by using localization. The second manifestation is higher-derivative contact interactions which arise from integrating out heavy string states. This part is more difficult and we can only work it out order by order in $1/\lambda$. At order $\mathcal{O}(\lambda^{-\frac{m}{2}})$, we expect up to $2m-2$ derivatives in the contact vertices \cite{Pufu:2023vwo}. Note that these higher-derivative corrections need to be compatible with superconformal symmetry. Therefore, a simple strategy is to first write down a general ansatz and use the superconformal Ward identities (\ref{scfWardid}) to fix it as much as possible. We then use localization and the flat-space limit to further constrain the remaining coefficients. Here we will analyze the solutions to the superconformal constraints which are the first step in this algorithm. 

Let us first consider higher-derivative corrections in position space. We can recursively use the the propagator identity which led to (\ref{Wcon2der}) to compute general higher-derivative contact Witten diagrams. It is not difficult to see that the following functions form a basis of contact Witten diagrams with no more than $2L$ derivatives
\begin{equation}\label{BMAbasis}
\{\mathcal{B}_{M,A}\}=\{\xi^A\chi^{M-A} \mathcal{W}^{\rm con}_{2+M,2+M}\}\;,\quad  L\geq M \geq A \in \mathbb{Z}_{\geq0}\;,
\end{equation}
where $\mathcal{W}^{\rm con}_{\Delta_1,\Delta_2}$ is the zero-derivative contact Witten diagram given in (\ref{Wcon0der}). Note these functions satisfy the weight-shifting differential relation \cite{Rigatos:2022eos}
\begin{equation}
\mathcal{B}_{M+1,A+1}=\frac{3+2M}{2(2+M)^2}(A\mathcal{B}_{M,A}-\xi\partial_\xi \mathcal{B}_{M,A})\;.
\end{equation}
For each $L$, we will write down the following ansatz
\begin{equation}
\mathcal{F}^{\rm h.d.}_L=\sum_{j=0}^2 \sum_{M=0}^L\sum_{A=0}^M c_{j,M,A} \sigma^j \mathcal{B}_{M,A}\;,
\end{equation}
where $c_{j,M,A}$ are unfixed coefficients. For $L=1$, we find that the superconformal Ward identities yield no solution. This is in agreement with the fact that (\ref{localIT}) has no correction at $\mathcal{O}(\lambda^{-1})$. For $L=2$, we find a unique solution up to an overall constant
\begin{equation}\label{AhdLeq2}
\begin{split}
\mathcal{F}^{\rm h.d.}_{L=2}={}&\frac{\mathcal{N}_{\rm sugra}^{\mathbb{T}}}{\pi}\frac{48c^{(2)}_1}{5}\bigg(\frac{7 \sigma ^2-14 \sigma +3}{12}\mathcal{B}_{0,0}+\frac{(64-29 \sigma)\sigma}{18}\mathcal{B}_{1,0}\\
{}&+\frac{-29 \sigma ^2+58 \sigma +6}{18}\mathcal{B}_{1,1}+\sigma^2 \mathcal{B}_{2,0}+2 (\sigma -1) \sigma\mathcal{B}_{2,1}+(\sigma -1)^2\mathcal{B}_{2,2}\bigg)\;.
\end{split}
\end{equation}
Similarly, we can work out the solutions for higher $L$. However, we find more unfixed coefficients in these solutions. For example, for $L=3$ we find 3 unfixed coefficients and for $L=4$ we find 6 unfixed coefficients. These solutions are more complicated in position space and we will not write them down here explicitly. 

Alternatively, we can translate the superconformal Ward identities (\ref{scfWardid}) into Mellin space following the strategy of \cite{Zhou:2017zaw}. The idea is to take the symmetric and antisymmetric combinations of (\ref{scfWardid}) to restore the invariance under $(z,\bar{z})\leftrightarrow(1/z,1/\bar{z})$, $(z,\bar{z})\leftrightarrow(\bar{z},z)$ obeyed by the Mellin representation \cite{Gimenez-Grau:2023fcy}. Then the terms in these combinations can written in terms of polynomials of $\xi$ and $\chi$, which can be interpreted as difference operators in Mellin space. Let us denote 
\begin{equation}
\mathbb{W}(z,\bar{z})=\left(\partial_z+\frac{1}{2}\partial_\alpha\right)\mathcal{F}(z,\bar{z};\alpha)\bigg|_{\alpha=z}\;,
\end{equation}
\begin{equation}
\mathbb{W}_+(z,\bar{z})=\mathbb{W}(z,\bar{z})+\mathbb{W}(1/z,1/\bar{z})\;,\quad \mathbb{W}_-(z,\bar{z})=\frac{1-z^2}{2z}\left(\mathbb{W}(z,\bar{z})-\mathbb{W}(1/z,1/\bar{z})\right)\;.
\end{equation}
Then we will consider the four indepndent linear combinations
\begin{equation}\label{Wpmpm}
\begin{split}
\mathbb{W}_{++}={}&\mathbb{W}_+(z,\bar{z})+\mathbb{W}_+(\bar{z},z)\;,\quad \mathbb{W}_{+-}=\frac{z\bar{z}}{(\bar{z}-z)(1-z\bar{z})}\left(\mathbb{W}_+(z,\bar{z})-\mathbb{W}_+(\bar{z},z)\right)\;,\\
\mathbb{W}_{-+}={}&\mathbb{W}_-(z,\bar{z})+\mathbb{W}_-(\bar{z},z)\;,\quad \mathbb{W}_{--}=\frac{z\bar{z}}{(\bar{z}-z)(1-z\bar{z})}\left(\mathbb{W}_-(z,\bar{z})-\mathbb{W}_-(\bar{z},z)\right)\;.
\end{split}
\end{equation}
Each of them will translate into a difference equation for the Mellin amplitude. The details of this translation can be found in Appendix \ref{App:scfWardMellin}. 

We can repeat the analysis in Mellin space where the ansatz is a polynomial with degree $L$ in $\delta$, $\gamma$ and degree 2 in $\sigma$. Again we find not solution at $L=1$ and the first solution appears at $L=2$. The position space solution (\ref{AhdLeq2}) corresponds to 
\begin{equation}
\mathcal{M}^{\rm h.d.}_{L=2}=\frac{\mathcal{N}_{\rm sugra}^{\mathbb{T}}}{\pi}c^{(2)}_1 M_{L=2}^{(1)}\;,
\end{equation}
where 
\begin{equation}\label{MLeq2}
M_{L=2}^{(1)}=\frac{1}{5} \left(5 \delta ^2+11 \delta +12-2 \sigma  (5 \gamma  \delta -32 \gamma +8 \delta +28)+(\gamma -2) (5 \gamma -14) \sigma ^2\right)\;,
\end{equation}
For $L=3$, the general solution can be written as 
\begin{equation}
\mathcal{M}^{\rm h.d.}_{L=3}=\frac{\mathcal{N}_{\rm sugra}^{\mathbb{T}}}{\pi}(c^{(3)}_1M_{L=3}^{(1)}+c^{(3)}_2M_{L=3}^{(2)}+c^{(3)}_3M_{L=2}^{(1)})\;,
\end{equation}
where the two new solutions are
\begin{equation}
\begin{split}\label{MLeq3}
M_{L=3}^{(1)}={}&\frac{1}{21}\bigg(21 \gamma  \delta ^2+51 \gamma  \delta +60 \gamma +8 \delta ^2+8 \delta-2 \sigma  \left(21 \gamma ^2 \delta -174 \gamma ^2+44 \gamma  \delta +244 \gamma +8 \delta -152\right)\\
{}&+(\gamma -2) \left(21 \gamma ^2-88 \gamma +76\right) \sigma ^2\bigg)\;,\\
M_{L=3}^{(2)}={}&\frac{1}{7}\bigg(\delta  (\delta +1) (7 \delta +4)-2 \sigma  \left(7 \gamma  \delta ^2-69 \gamma  \delta +100 \gamma +18 \delta ^2+14 \delta -80\right)\\
{}&+(\gamma -2) \sigma ^2 (7 \gamma  \delta -16 \gamma -20 \delta +44)\bigg)\;.
\end{split}
\end{equation}
To connect them with the basis in (\ref{BMAbasis}), we note each function has polynomial Mellin amplitude of degree $M$
\begin{equation}\label{MellinBMA}
\mathcal{M}[\mathcal{B}_{M,A}]=\frac{2\Gamma(\frac{3}{2}+M)}{\sqrt{\pi}\Gamma(2+M)^2}(\gamma-\delta)_{M-A}(\delta)_A\;.
\end{equation}
We also note that in the large $\delta$, $\gamma$ limit
\begin{equation}
\begin{split}
M_{L=2}^{(1)}(\Lambda \delta,\Lambda\gamma)={}&\Lambda ^2 (\delta -\gamma  \sigma )^2+\ldots\;,\\
M_{L=3}^{(1)}(\Lambda \delta,\Lambda\gamma)={}&\Lambda ^3 (\delta -\gamma  \sigma )^2\gamma+\ldots\;,\\
M_{L=3}^{(2)}(\Lambda \delta,\Lambda\gamma)={}&\Lambda ^3 (\delta -\gamma  \sigma )^2\delta+\ldots\;,
\end{split}
\end{equation}as we take $\Lambda\to\infty$. For $L=4$, we find three more solutions
\begin{equation}
\mathcal{M}^{\rm h.d.}_{L=4}=\frac{\mathcal{N}_{\rm sugra}^{\mathbb{T}}}{\pi}(c^{(4)}_1M_{L=4}^{(1)}+c^{(4)}_2M_{L=4}^{(2)}+c^{(4)}_3M_{L=4}^{(3)}+c^{(4)}_4M_{L=3}^{(1)}+c^{(4)}_5M_{L=3}^{(2)}+c^{(4)}_6M_{L=2}^{(1)})\;,
\end{equation}
where the new solutions have the leading behavior
\begin{equation}
\begin{split}
M_{L=4}^{(1)}(\Lambda \delta,\Lambda\gamma)={}&\Lambda ^4 (\delta -\gamma  \sigma )^2\gamma^2+\ldots\;,\\
M_{L=4}^{(2)}(\Lambda \delta,\Lambda\gamma)={}&\Lambda ^4 (\delta -\gamma  \sigma )^2\gamma\delta+\ldots\;,\\
M_{L=4}^{(3)}(\Lambda \delta,\Lambda\gamma)={}&\Lambda ^4 (\delta -\gamma  \sigma )^2\delta^2+\ldots\;,
\end{split}
\end{equation}
and are given explicitly in Appendix \ref{App:scfWardMellin}. In general, for stringy corrections with no more than $2L$ derivatives the superconformal Ward identities fix the ansatz to be the linear combination of $\frac{1}{2}L(L-1)$ solutions. In the large energy limit, these solutions span degree $L-2$ polynomials of $\delta$, $\gamma$ after exacting a factor $(\delta -\gamma  \sigma )^2$. Once we have found these solutions in Mellin space, they can be easily translated back into position space by using (\ref{MellinBMA}). 

\subsubsection{Stringy corrections up to eight derivatives}
We now use constraints from supersymmetric localization and the flat-space limit to study the stringy corrections up to $L=4$. As mentioned, the position space correlator can be expanded as 
\begin{equation}
\mathcal{F}=\mathcal{F}_{\rm sugra} f(\lambda)+\lambda^{-1}\mathcal{F}^{\rm h.d.}_{L=2}+\lambda^{-\frac{3}{2}}\mathcal{F}^{\rm h.d.}_{L=3}+\lambda^{-2}\mathcal{F}^{\rm h.d.}_{L=4}+\ldots\;.
\end{equation}
Here $f(\lambda)$ is the one-point function coefficient normalized by its leading term at large $\lambda$
\begin{equation}
f(\lambda)=\frac{a_S(\lambda)}{a_S(\lambda)|_{\rm leading}}=\sqrt{1+\frac{\pi^2}{\lambda}}\;.
\end{equation}
where we have used (\ref{aL}) to get
\begin{equation}
a_S(\lambda)=-\frac{\pi\sqrt{2(\lambda+\pi^2)}}{\lambda}+\mathcal{O}(1/N)\;.
\end{equation}
Correspondingly, we have in Mellin space 
\begin{equation}
\mathcal{M}=\mathcal{M}_{\rm sugra} f(\lambda)+\lambda^{-1}\mathcal{M}^{\rm h.d.}_{L=2}+\lambda^{-\frac{3}{2}}\mathcal{M}^{\rm h.d.}_{L=3}+\lambda^{-2}\mathcal{M}^{\rm h.d.}_{L=4}+\ldots\;.
\end{equation}
It is straightforward to integrate the ansatz and we get the expansion
\begin{equation}\label{Fhdint}
\begin{split}
{}&\int_1^\infty d\zeta\int_{-1}^1d\eta \mathcal{F}_2(\zeta,\eta)=\mathcal{N}_{\rm sugra}^{\mathbb{T}}\bigg(-\frac{1}{4}+\frac{1}{\lambda}\bigg(\frac{8c_1^{(2)}-5\pi^2}{40}\bigg)+\frac{1}{\lambda^{\frac{3}{2}}}\bigg(\frac{14c_1^{(3)}-6c_2^{(3)}+21c_3^{(3)}}{105}\bigg)\\
{}&\quad\quad+\frac{1}{\lambda^2}\bigg(\frac{1575\pi^4-32(76c_1^{(4)}+5(4c_2^{(4)}+11c_3^{(4)}-42c_4^{(4)}+18c_5^{(4)}-63c_6^{(4)}))}{50400}\bigg)+\ldots\bigg)\;.
\end{split}
\end{equation}
This will be compared with the integrated correlator (\ref{localIT}). On the other hand, the twisted correlator reads
\begin{equation}\label{Fhdtwist}
\begin{split}
\mathcal{F}_{\rm top}(-1)={}&\mathcal{N}_{\rm sugra}^{\mathbb{T}}\bigg(\frac{3}{8}+\frac{1}{\lambda}\bigg(\frac{15\pi^2+8c_1^{(2)}}{80}\bigg)+\frac{1}{\lambda^{\frac{3}{2}}}\bigg(\frac{20c_1^{(3)}+21c_3^{(3)}}{210}\bigg)\\
{}&+\frac{1}{\lambda^2}\bigg(\frac{-2835\pi^4+32(56c_1^{(4)}+180c_4^{(4)}+189c_6^{(4)})}{60480}\bigg)+\ldots\bigg)\;,
\end{split}
\end{equation}
which is to be compared with the localization result (\ref{localET}). The flat-space amplitude (\ref{Adiskorth}) in the orthogonal configuration (\ref{orthcondi}) can be expanded in powers of $\ell_s$, or equivalently in $1/\lambda=\ell_s^4/R^4$, as 
\begin{equation}
\begin{split}
(u_1\cdot \theta)^{-2}(u_2\cdot \theta)^{-2}\mathcal{A}^{\rm disk}_{h\to h}={}&-T_{\rm D1}(S+Q\sigma)^2\bigg(\frac{1}{QS}-\frac{\pi^2R^4}{12\lambda}-\frac{(2Q+S)\zeta(3)R^6}{4\lambda^{\frac{3}{2}}}\\
{}&\quad\quad-\frac{(8Q^2+QS+2S^2)\pi^4R^8}{1440\lambda^2}+\ldots\bigg)\;,
\end{split}
\end{equation}
which translates via (\ref{fslformula}) to the following large $\delta$, $\gamma$ behavior of the Mellin amplitude
\begin{equation}\label{Mhe}
\mathcal{M}(\delta,\gamma)\approx \frac{\mathcal{N}_{\rm sugra}^{\mathbb{T}}(\delta-\gamma\sigma)^2}{\pi} \left(\frac{1}{\gamma\delta}+\frac{5\pi^2}{4\lambda}+\frac{105(2\gamma-\delta)\zeta(3)}{4\lambda^{\frac{3}{2}}}+\frac{21\pi^4(8\gamma^2-\gamma\delta+2\delta^2)}{32\lambda^2}+\ldots\right)\;.
\end{equation}
Note that in the large $R$ limit, only the terms with the highest scaling power at each order in $1/\lambda$ will survive in the flat-space formula. 

Let us now consider the stringy corrections at $L=2$. From (\ref{localIT}), (\ref{localET}) and the flat-space limit (\ref{Mhe}) we have three conditions for just one number. We find they all three fix the coefficient to be 
\begin{equation}
c_1^{(2)}=\frac{5\pi^2}{4}\;,
\end{equation}
providing a nontrivial consistency check of our method. At the next order $L=3$, the flat-space limit gives
\begin{equation}
c_1^{(3)}=\frac{105}{2}\zeta(3)\;,\quad c_2^{(3)}=-\frac{105}{4}\zeta(3)\;.
\end{equation}
The remaining coefficient $c_3^{(2)}$ is constrained by the two localization conditions. Nontrivially, they are solved by
\begin{equation}
c_3^{(3)}=-50\zeta(3)\;,
\end{equation}
which provides a further consistency check. For $L=4$, the flat-space limit fixes the three leading coefficients
 \begin{equation}
c_1^{(4)}=\frac{21\pi^4}{4}\;,\quad c_2^{(4)}=-\frac{21\pi^4}{32}\;,\quad c_3^{(4)}=\frac{21\pi^4}{16}\;.
\end{equation}
However, we have only two conditions for three unfixed coefficients
\begin{equation}
14 c_4^{(4)}-6c_5^{(4)}+21 c_6^{(4)}=\frac{2093\pi^4}{80}\;,\quad 20 c_4^{(4)}+21 c_6^{(4)}=-\frac{1099\pi^4}{24}\;.
\end{equation}
Therefore, we would need another condition in order to fully fix stringy corrections at this order.  

\section{Discussions and outlook}\label{Sec:outlook}
In this paper, we proposed a flat-space limit formula for holographic defect correlators and checked its validity in several nontrivial examples. We then applied this formula to study the two-point function in $\mathcal{N}=4$ SYM in the presence of a 't Hooft loop. Together with constraints from supersymmetric localization, we fixed all higher-derivative stringy corrections with eight derivatives or fewer up to a linear combination. Our results lead to an array of future research directions and we briefly outline some of them below. 

In this paper, we mainly focused on defect two-point functions for the simplicity. However, similar arguments also apply for the flat-space limit of higher-point correlators where operators can be inserted both in the bulk and on the defect. Our formula (\ref{fslformula}) extends to the general case with the obvious modifications where $\Delta_1+\Delta_2$ becomes the sum of dimensions, etc. Note that in particular when all operators are inserted on the defect, the formula (\ref{fslformula}) reduces to the defect-free formula (\ref{flatspacePenedones}) where we set $d=p$.  Moreover, we considered the generic case where the co-dimension of the defect $p$ is larger than 1. The generalization to the BCFT case with $p=1$ is straightforward and the flat-space formula should be completely analogous in the BCFT Mellin space formalism \cite{Rastelli:2017ecj}. On the other hand, we only give the  prescription for taking the flat-space limit in Mellin space. Starting from this result, it would also be interesting to know how to take this limit in position space, along the lines of \cite{Maldacena:2015iua} for the defect-free case. 

As a concrete example, we considered in detail the case of 't Hooft loops in $\mathcal{N}=4$ SYM. However, there are other types of line defects in this theory which are dual to general $(p,q)$ strings in AdS and are related to each other by $SL(2,\mathbb{Z})$ transformations. Integrated correlators for these line defects have been recently discussed in \cite{Pufu:2023vwo,Billo:2023ncz,Dempsey:2024vkf,Billo:2024kri,Dorigoni:2024vrb,Dorigoni:2024csx} and these results encode nontrivial information about the stringy corrections. It would be interesting to extend the bootstrap analysis performed in this paper to general Wilson-'t Hooft lines by combining localization and the flat-space limit. More generally, we can perform similar analyses in other theories and study correlators of local operators in the presence of defects of other dimensions. 

Finally, for illustrative purposes we restricted ourselves to a very basic approach to stringy corrections in this paper. We simply enumerated the possible superconformal solutions and then used constraints from localization and flat-space limit to fix their coefficients. Such an approach suffers from the drawback that we will have increasingly more unfixed coefficients at higher orders as the flat-space limit can only determine the leading coefficients and the information provided by localization is finite. In the defect-free case, there is also a more sophisticated approach to
study stringy corrections in four-point functions
where worldsheet intuition and single-valuedness allow for infinitely many coefficients to be fixed  \cite{Alday:2022uxp,Alday:2022xwz,Alday:2023jdk,Alday:2023mvu}.  It would be very interesting to explore a similar worldsheet method for defect two-point functions and compute the AdS string form factor in a $1/R$ expansion.

\acknowledgments
The work of L.F.A. is partially supported by the STFC grant ST/T000864/1. L.F.A. would like to thank ShanghaiTech University for their hospitality during completion of this work. 
The work of X.Z. is supported by the NSFC Grant No. 12275273, funds from Chinese Academy of Sciences, University of Chinese Academy of Sciences, and the Kavli Institute for Theoretical Sciences. The work of X.Z. is also supported by the Xiaomi Foundation.

\appendix

\section{More details of the check at one loop}\label{App:1loopcheck}
Here we give more details about how to take the large $m$, $n$ limit of $c_{mn}$ in (\ref{M1loop}). The explicit expression of $c_{mn}$ was given in \cite{Chen:2024orp} in terms of the following hypergeometric function
\begin{equation}
h(m,n)= \frac{\sqrt{\pi }4^{ m}\Gamma (m+n+3)}{\Gamma(\tfrac{-2m-3}{2}) \Gamma (2 m+n+6)}{}_3 F_2\left(\left.\begin{gathered}
-m-2\,,\tfrac{-2m-n-5}{2}\,, \tfrac{-2m-n-4}{2} \\
\tfrac{-2m-3}{2}\,, -m-n-2
\end{gathered} \right\rvert\, 1\right)\;.
\end{equation}
First we note that $h(m,n)$ satisfies the following simple recursion relation
\begin{equation}
(2m-n+5)h(m,n)+2(n+1)h(m,n+1)-(2m+n+7)h(m+1,n)=0\;.
\end{equation}
We look for solutions to this equation in the large $m$, $n$ limit with fixed $m/n$. To do this, we can introduce $m=\Lambda x$, $n=\Lambda y$ and take $\Lambda\to\infty$ with $x$, $y$ fixed. The recursion relation becomes 
\begin{equation}
(2\Lambda x-\Lambda y+5)h(x,y)+2(\Lambda y+1)h(x,y+1/\Lambda)-(2\Lambda x+\Lambda y+7)h(x+1/\Lambda,y)=0\;,
\end{equation}
which gives a set of differential equations in a large $\Lambda$ expansion relating coefficient functions in $x$, $y$ at different orders in $1/\Lambda$. These differential equations, are further supplemented by the following information from small values $m$. For fixed $m$, such as $m=1,2,\cdots$, we can explicitly find $h(m,n)$ as a function of $n$, for example
\begin{equation}
h(1,n)=\frac{1}{16}\left(\frac{1}{n+1}-\frac{6}{n+2}+\frac{24}{n+3}-\frac{56}{n+4}+\frac{78}{n+5}-\frac{60}{n+6}+\frac{20}{n+7}\right)\;.
\end{equation}
It is not difficult to find the following uniform large $n$, finite $m$, behavior
\begin{equation}
h(m,n)=\frac{1}{16n}-\frac{\frac{m}{8}+\frac{5}{16}}{n^2}+\frac{\frac{3m^2}{8}+\frac{7m}{4}+\frac{33}{16}}{n^3}+\ldots\;.
\end{equation}
When we set $m=\Lambda x$, $n=\Lambda y$ and take $\Lambda\to\infty$, we find that the leading term in $m$ at each order in $1/n$ contributes in this limit while subleading terms will contribute to higher orders in $1/\Lambda$. It is not difficult to resum these contributions and we find 
\begin{equation}
h(m,n)=\frac{1}{16\sqrt{n(4m+n)}}\left(1+\frac{1}{n}h_1(n/m)+\ldots\right)\;,
\end{equation}
where $\ldots$ are higher orders in $1/\Lambda$ and
\begin{equation}
h_1(z)=-\frac{z(5z+22)+2}{(z+4)^2}\;.
\end{equation}
It is easy to check that it satisfies the differential equations arising from the large $\Lambda$ expansion of the recursion relation. Using this expansion in $c_{mn}$, we find for large $m$, $n$
\begin{equation}
c_{mn}\approx\frac{15m^2n^{\frac{3}{2}}}{4(4m+n)^{\frac{5}{2}}}\;.
\end{equation}

\section{Laguerre polynomial asymptotics}\label{App:LaguerrelargeN}
In computing the coupling derivative (\ref{defEL}), we encounter ratios of Laguerre polynomials of the following two types
\begin{equation}
R_1(x)=\frac{L_{N-2}^2(-Nx)}{L_{N-1}^1(-Nx)}\;,\quad R_2(x)=\frac{L_{N-2}^3(-Nx)}{L_{N-1}^1(-Nx)}\;.
\end{equation}
We need to take the large $N$ limit where $x>0$ is kept fixed. To get the asymptotics of these ratios, we first notice that they satisfy the following identity as a consequence of the recurrence relations for Laguerre polynomials
\begin{equation}
xR_2(x)+\left(\frac{2}{N}+x\right)R_1(x)+\frac{1-N}{N}=0\;.
\end{equation}
Furthermore, by taking derivatives and using the definition of Laguerre polynomials we find
\begin{equation}
\frac{dR_1(x)}{dx}=NR_2(x)-NR_1^2(x)\;.
\end{equation}
Combining these two identities, we find a nonlinear differential equation for $R_1(x)$
\begin{equation}
\frac{dR_1(x)}{dx}+\frac{N(xR_1(x)^2+xR_1(x)-1)+2R_1(x)+1}{x}=0\;.
\end{equation}
We can use this differential equation to take the large $N$ limit of $R_1(x)$ with fixed $x$ and we find
\begin{equation}
R_1(x)=\frac{\sqrt{x(x+4)}-x}{2x}-\frac{1}{N}\frac{x+3}{x(x+4)}+\ldots\;.
\end{equation}

\section{Superconformal Ward identities in Mellin space}\label{App:scfWardMellin}
We consider more generally the correlator $\llangle S_{p_1}S_{p_2}\rrangle$ of two $\frac{1}{2}$-BPS operators with dimensions $p_1$, $p_2$. We expand the two-point function into powers of the R-symmetry cross ratio
\begin{equation}
\mathcal{F}(\xi,\chi;\sigma)=\sum_{j=0}^{p_m}\sigma^j\mathcal{F}_j(\xi,\chi)\;,
\end{equation}
where $p_m=\min\{p_1,p_2\}$. The superconformal Ward identity becomes (\ref{scfWardid})
\begin{equation}
\mathbb{W}(z,\bar{z})=\sum_{j=0}^{p_m}\frac{(1-z)^{2j}}{z^j}\left(\left(\frac{\bar{z}-1}{\sqrt{z\bar{z}}}-\frac{\xi}{2z}\right)\partial_\xi+\left(\frac{1}{\sqrt{z\bar{z}}}-\frac{\chi}{2z}\right)\partial_\chi+\frac{j(z+1)}{2z(z-1)}\right)\frac{\mathcal{F}_j(\xi,\chi)}{(-2)^j}\;.
\end{equation}
Then we have
\begin{equation}
\begin{split}
\mathbb{W}_{+}(z,\bar{z})={}&\sum_{j=0}^{p_m}\frac{(1-z)^{2j}}{z^j}\bigg(-\left(\frac{(1-z)^2}{2z}+2\right)\xi\partial_\xi+\left(\xi-\frac{(1-z)^2}{2z}\chi\right)\partial_\chi\\
{}&\quad\quad-\frac{j(1-z)^2}{2z}-2j\bigg)\frac{\mathcal{F}_j(\xi,\chi)}{(-2)^j}\;,\\
\mathbb{W}_{-}(z,\bar{z})={}&\sum_{j=0}^{p_m}\frac{(1-z)^{2j}}{z^j}\bigg(-\left(\left(\frac{(1-z)^2}{2z}\right)^2+\frac{3(1-z)^2}{2z}+2\right)\xi\partial_\xi\\
{}&\quad\quad+\left(\left(\frac{(1-z)^2}{2z}+1\right)\xi-\left(\left(\frac{(1-z)^2}{2z}\right)^2+\frac{(1-z)^2}{2z}\right)\chi\right)\partial_\chi\\
{}&\quad\quad-j\left(\frac{(1-z)^2}{2z}\right)^2-\frac{3j(1-z)^2}{2z}-2j\bigg)\frac{\mathcal{F}_j(\xi,\chi)}{(-2)^j}\;.
\end{split}
\end{equation}
Upon further symmetrizing and antisymmetrizing, we encounter the following combinations of cross ratios
\begin{equation}
\zeta^{(j)}_+=\frac{(1-z)^{2j}}{2z^j}+\frac{(1-\bar{z})^{2j}}{2\bar{z}^j}\;,\quad \zeta^{(j)}_-=\frac{z\bar{z}}{(\bar{z}-z)(1-z\bar{z})}\left(\frac{(1-z)^{2j}}{2z^j}-\frac{(1-\bar{z})^{2j}}{2\bar{z}^j}\right)\;,
\end{equation}
which, as was pointed out in \cite{Gimenez-Grau:2023fcy}, can be expressed as polynomials of $\xi$ and $\chi$. In Mellin space, we have the following simple translation 
\begin{equation}
\begin{split}
{}&\xi\partial_\xi\to(-\delta)\times\;,\quad \chi\partial_\chi\to(-\gamma+\delta)\times\;,\\
{}&\xi^m\chi^n\to \widehat{\xi^m\chi^n}\circ \mathcal{M}(\delta,\gamma)=\mathcal{M}(\delta+m,\gamma+m+n)(\delta)_m(\gamma-\delta)_n\prod_{i=1}^2\left(\frac{\Delta_i-\gamma}{2}\right)_{-\frac{m+n}{2}}\;.
\end{split}
\end{equation}
Translating term by term, we get four difference equations from (\ref{Wpmpm}).  

Using these equations, it is not difficult to find the solutions (\ref{MLeq2}) and (\ref{MLeq3}) for $L$ up to 3. For $L=4$, we find 
\begin{equation}
\begin{split}\label{MLeq4}
M_{L=4}^{(1)}={}&\frac{1}{135}\bigg(15 \gamma ^2 \left(9 \delta ^2+23 \delta +28\right)+72 \gamma  \delta  (\delta +1)-28 \delta ^2 (\delta +1)\\
{}&+\sigma  \left(-30 \gamma ^3 (9 \delta -92)-48 \gamma ^2 (13 \delta +123)+8 \gamma  \left(7 \delta ^2-91 \delta +1028\right)+16 \left(9 \delta ^2+5 \delta -242\right)\right)\\
{}&(\gamma -2) \sigma ^2 \left(135 \gamma ^3-798 \gamma ^2-4 \gamma  (7 \delta -377)+80 \delta -984\right)\bigg)\;,\\
M_{L=4}^{(2)}={}&\frac{1}{9}\bigg(\delta  (\delta +1) (\gamma  (9 \delta +4)-4 \delta )+(\gamma -2)^2 \sigma ^2 (3 \gamma  (3 \delta -8)-28 \delta +76)\\
{}&+\sigma  \left(-2 \gamma ^2 \left(9 \delta ^2-109 \delta +196\right)-8 \gamma  \left(5 \delta ^2+34 \delta -104\right)+24 \left(\delta ^2+7 \delta -24\right)\right)\bigg)\;,\\
M_{L=4}^{(3)}={}&\frac{1}{18}\bigg(3 \delta ^2 \left(6 \delta ^2+13 \delta +7\right)+\sigma  \left(-6 \gamma  \left(6 \delta ^3-73 \delta ^2+165 \delta -100\right)-128 \delta ^3+92 \delta ^2+484 \delta -480\right)\\
{}&+\sigma ^2 \left(3 \gamma ^2 \left(6 \delta ^2-21 \delta +16\right)+\gamma  \left(-88 \delta ^2+302 \delta -228\right)+8 \left(13 \delta ^2-44 \delta +33\right)\right)\bigg)\;.
\end{split}
\end{equation}

\bibliography{refs} 
\bibliographystyle{utphys}
\end{document}